\documentclass[journal,twoside,web]{ieeecolor}
\bstctlcite{IEEEexample:BSTcontrol}
\usepackage{generic}
\usepackage{cite}
\usepackage{amsmath,amssymb,amsfonts}
\usepackage{algorithmic}
\usepackage{graphicx}
\usepackage{textcomp}
\usepackage[caption=false]{subfig}
\def\BibTeX{{\rm B\kern-.05em{\sc i\kern-.025em b}\kern-.08em
		T\kern-.1667em\lower.7ex\hbox{E}\kern-.125emX}}
\begin{document}
	\title{Thickness dependence of space-charge-limited current in spatially disordered organic semiconductors}
	\author{Muhammad Zubair, \textit{Member, IEEE}, Yee Sin Ang, and Lay Kee Ang, \textit{Senior Member, IEEE}
		\thanks{Muhammad Zubair is with the Department of Electrical Engineering at Information Technology University of the Punjab, Ferozpur Road, 54000 Lahore, Pakistan. He was affiliated until recently with the SUTD-MIT International Design Center, Singapore University of Technology and Design, 487372 Singapore (e-mail: muhammad.zubair@itu.edu.pk).}
		\thanks{Yee Sin Ang, and Lay Kee Ang are affiliated with the SUTD-MIT International Design Center, Singapore University of Technology and Design, 487372 Singapore (e-mails: yee\_sin@sutd.edu.sg, ricky\_ang@sutd.edu.sg). }	
		\thanks{The authors are thankful to P. W. M. Blom for providing the experimental data of Ref.~\cite{blom2005thickness} and helpful
			discussion. This work is sponsored by USA AFOSR AOARD (FA2386-14-1-4020), Singapore Temasek Laboratories (IGDS S16 02 05 1) and A*STAR IRG (A1783c0011).}
	}
	\maketitle

\begin{abstract}
Charge transport properties in organic semiconductors are determined
by two kinds of microscopic disorders, namely energetic disorder
related to the distribution of localized states and the spatial disorder
related to the morphological features of the material. From a semi-classical picture, the charge transport properties are crucially determined by both the carrier mobility and the electrostatic field distribution in the material. Although the effect of disorders on carrier mobility has been widely studied, how electrostatic field distribution is distorted by the presence of disorders and its effect on charge transport remain unanswered. In this paper, we present a modified space-charge-limited current (SCLC) model for spatially disordered organic semiconductors based on the fractional-dimensional electrostatic framework. We show that the thickness dependence of SCLC is related to the spatial disorder in organic semiconductors. 
For trap-free transport, the SCLC exhibits a modified thickness scaling of $J\propto L^{-3\alpha}$, where the fractional-dimension parameter $\alpha$ accounts for the spatial disorder in organic semiconductors. The trap-limited and field-dependent mobility are also shown to obey an $\alpha$-dependent thickness scaling. The modified SCLC model shows a good agreement with several experiments on spatially disordered organic semiconductors. By applying this model to the experimental data, the standard charge transport parameters can be deduced with better accuracy than by using existing models.




\end{abstract}

\maketitle


\section{\label{sec:Introduction} Introduction}

The mobility of charge carrier is a key parameter for the performance of optoelectronic devices~\cite{kuik201425th}, especially for devices using organic semiconductors and polymers.
The mobility in organic semiconductors strongly depends on the nature, structure, purity of the materials and device operating conditions.
The charge transport in organic compounds occurs across various levels, ranging from within molecules, between molecules as well as between crystalline grains and amorphous and crystalline regions.
The transport properties are determined by two kinds of microscopic disorders, namely the energetic disorder characterized by a broad distribution of localized states and the spatial disorder related to the morphological features of the material~\cite{campbell2016charge}. The space-charge-limited current (SCLC) is an important classical transport phenomenon in organic semiconductors where the quantum effects can be ignored at microscale and above~\cite{kuik201425th, campbell2016charge, lampert1970current, mott1948electronic,mark1962space,murgatroyd1970theory,bassler1993charge,blakesley2014towards,tanase2003unification,blom2005thickness,fishchuk2007analytic,pasveer2005unified,cottaar2011scaling,tanase2004charge, fishchuk2010temperature,katsouras2013charge,leijtens2013charge}. 

The mobility of a given organic material sandwiched between two
planar electrodes with an applied voltage ($V_0$), is commonly
measured indirectly by fitting the measured current density-voltage
(J-V) characteristics at high voltages to some SCLC models~\cite{lampert1970current}.  It is assumed that there is no barrier (ohmic contact) at the interface
when the charges are injected from the electrode into the solid. The
simplest SCLC model for a trap-free solid is known as the
one-dimensional (1D) classical Mott-Gurney (MG) law
~\cite{mott1948electronic}, given by

\begin{eqnarray}
J= \frac{9}{8}\epsilon \mu\frac{V_{0}^{2}}{L^{3}},
\label{eqn:MGClassical}
\end{eqnarray}
where $\epsilon=\epsilon_{0}\epsilon_{r}$ is permittivity, and $L$ is the thickness of the solid embedded between the metal electrodes.
Once the values of $J$, $V_0$, $L$ and $\epsilon$ are determined, the mobility $\mu$ can be calculated by fitting the J-V characteristics at high $V_0$ region to the 1D MG law.
Note the assumptions in using the MG law include constant mobility (independent of the applied electric field and charge density) and the solid is trap-free.

For a trap-filled solid with exponentially energy-distributed traps, the corresponding SCLC model is known as the trap-limited (TL) SCLC model ~\cite{mark1962space}:
\begin{eqnarray}
J= N_{c} \mu e^{1-l} \left[ \frac{\epsilon l}{N_{t}(l+1)}\right]^{l}\left( \frac{2l+1}{l+1} \right) ^{l+1}\frac{V_{0}^{l+1}}{L^{2l+1}},
\label{eqn:Tl-SCLC-Classical}
\end{eqnarray}
where $N_{c}$ is the effective density of states corresponding to the energy at the bottom of the conduction band, $N_{t}$ is the total trapped electron density, and $l=T_{t}/T \geq 1$ with $T_{t}$ being a parameter controlling the trap distribution.

For shallow traps or energetic disorder, the mobility varies with
the electric field $E$ that Murgatroyd's~\cite{murgatroyd1970theory}
model may be used to describe a field-dependent mobility in the form
of
\begin{eqnarray}
\mu=\mu_{0} \exp(\gamma \sqrt{E}),
\label{eqn:field_dependence}
\end{eqnarray}
where $\mu_{0}$ is representing the mobility at zero field, and
$\gamma$ is a material-specific parameter that describes the
strength of the field-dependence. The field-dependent mobility can also
include other effects, such as carrier-density dependence
(Gaussian disorder model (GDM)~\cite{bassler1993charge}) and deep
traps, which can only be resolved by having a more comprehensive
model to fit with experiments over a wide range of parameters
~\cite{blakesley2014towards}. By using Eq.
(\ref{eqn:field_dependence}), we get \cite{murgatroyd1970theory}
\begin{eqnarray}
J= \frac{9}{8}\epsilon \mu_{0}\frac{V_{0}^{2}}{L^{3}}\exp(0.89
\gamma \sqrt{E}), \label{eqn:MGClassical_fielddep}
\end{eqnarray}
where $0.89\gamma$ is used as a fitting parameter.

Other than field-dependent mobility, charge-carrier-density
dependent mobility models have also been studied including a
power-law dependence for energetically disordered semiconductors by
Tanase et. al.~\cite{tanase2003unification}, Blom et.
al.~\cite{blom2005thickness}, and others for disordered polymers
~\cite{fishchuk2007analytic,pasveer2005unified,cottaar2011scaling}.

The mobility is often extracted from J-V measurements by fitting the experimental data with the theoretical models of SCLC with different mobility terms. The mobility of a given sample is determined solely based on the goodness of the fit. Such empirical methodology may not always produce accurate physical picture. For example, an inconsistency of \textit{field-dependent} mobility model with the experimental data has been raised in Ref.~\cite{blom2005thickness}. The conventional model of Eq. ({\ref{eqn:MGClassical_fielddep}}) could not fit the experimental data due to weaker thickness dependence of measured SCLC, and hence the \textit{carrier-density dependent} mobility model was chosen to extract the mobility of PPV based diode device. However, assuming Gaussian density of states (DOS), it is known that in low carrier-density regime the mobility should be \textit{carrier-density independent}~\cite{tanase2004charge}. Another recent experiment~\cite{campbell2016charge} has also reported that the charge transport in amorphous semiconductors is \textit{not} charge density dependent but instead follows a field-dependent mobility model. In such scenarios, a new model of space-charge-limited transport is required to capture the correct thickness scaling of measured SCLC.

The transport sites in organic semiconductors are distributed both in space and energy. The combined effect of spatial and energetic disorder on charge transport has already been studied in the previous works (see~\cite{nenashev2015theoretical} for a comprehensive review). However, the SCLC is a transport phenomenon closely related with both the material properties (i.e., mobility) and the electrostatic field distribution inside these materials, and thus far, the non-uniform distribution of electrostatic field due to spatial complexity of the material and its implications on the macroscopic SCLC have not been fully addressed. The complex, spatially disordered and often self-organized microstructure, in which ordered microcrystalline domains are embedded in an amorphous domain, can be considered as fractal features having important consequences for electrical properties of these materials (refer to Fig. 2 in ~\cite{tessler2009charge} and Fig. 1 in~\cite{noriega2013general}). 

\begin{figure}[!hb]
	\centering
	\includegraphics[width=0.45\textwidth]{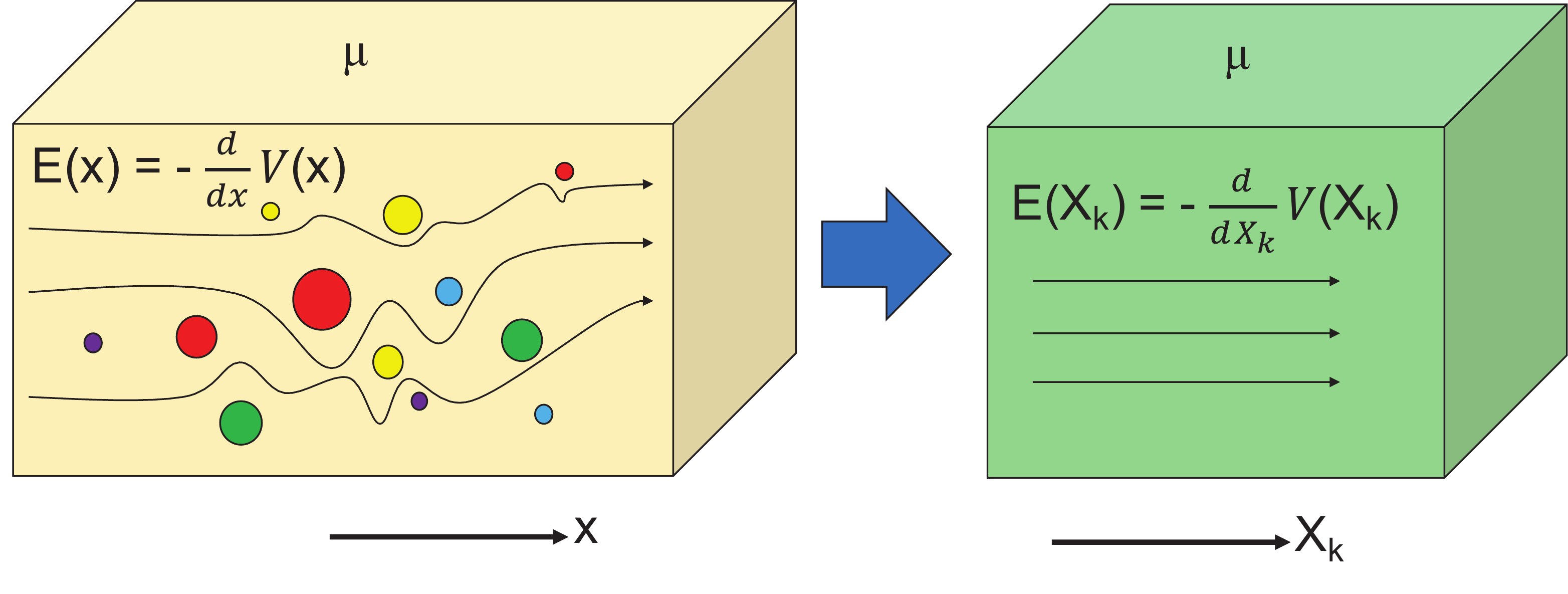}
	\caption{Schematic description of the concept that the \textit{spatially disordered} organic semiconductor in real space [shown on left] can be considered effectively as \textit{spatially ordered} organic semiconductor embedded in the corresponding fractional-dimensional space [shown on right] using fractional-dimensional space framework as description of complexity described in Appendix. $X_{k}$ corresponds to the spatial coordinates of an $\alpha$-dimensional space. }
	\label{fig:concept}
\end{figure}

In this work, we present a modified SCLC model to account for the 
	spatial disorder effect of a solid such as amorphous semiconducting
	polymer by treating the material as a fractal object. The key novelty of our proposed model is to utilize the
	fractional-dimension of space to effectively model the non-uniform distribution of electrostatic field inside these spatially disordered materials. Fig. (\ref{fig:concept}) provides a schematic description of the concept that the \textit{spatially disordered} organic semiconductor in real integer-dimensional space can be considered effectively as \textit{spatially ordered} organic semiconductor embedded in the corresponding fractional-dimensional space using mathematical framework briefly introduced in the Appendix (also see~\cite{zubair2017fractional,zubair2016fractional} and references therein).
	Such methods have been applied in other areas
	including quantum field
	theory~\cite{stillinger1977axiomatic,palmer2004equations},
	general relativity~\cite{sadallah2009solution},
	thermodynamics~\cite{tarasov2016heat},
	mechanics~\cite{ostoja2014fractal},
	hydrodynamics~\cite{balankin2012map},
	electrodynamics~\cite{zubair2012electromagnetic,mughal2011fractional,naqvi2016cylindrical, zubair2011exact, asad2012electromagnetic, asad2012reflection,zubair2011exact2,zubair2011differential,zubair2011electromagnetic,zubair2010wave
		}, and
	fractional charge transport~\cite{zubair2016fractional,zubair2017fractional} to name a few.

The proposed approach has been generalized to cover three types of SCLC models: trap-free  model (or MG law), trap-limited (TL) model and the field dependent mobility models. We first analyze various experimental
results to study the thickness (or $L$) dependence to show that the
traditional $L$ scalings from the traditional models are not valid
for spatially disordered semiconducting materials. By using our
proposed models, we are able to
reproduce the experimental current-voltage measurements in~\cite{blom2005thickness} without
using the carrier-density dependent model, and thus solving the issues
raised by recent paper~\cite{campbell2016charge}. 

\section{\label{sec: Formulation} Derivation of SCLC Scaling Laws for spatially disordered organic semiconductors}
\subsection{Trap-free model}
Here, we derive the modified MG law for spatially disordered organic semiconductors with the assumption that the effect of non-uniform electrostatic field distribution inside spatially disordered material can be studied effectively by replacing the governing equations of classical SCLC model with the fractional-dimensional counterparts, by using the formulation described in Appendix (for more details see~\cite{zubair2017fractional,zubair2016fractional} and references therein), where the fractional-dimension is related to the amount of spatial disorder. Provided that the thermal carriers are negligible in comparison to injected carrier, and assuming that the size of the electrode is much greater than the spacing $L$, thus the derivation is conducted only in $x$ direction perpendicular to surface of the electrode.
The following equations are  solved in the $\alpha$-dimensional space ~\cite{tarasov2014anisotropic} with $0<\alpha\leq1$:

\begin{eqnarray}
J=-\rho\nu \label{eqn:J=rho nu},\\
\nabla_{\alpha} \cdot E= \frac{1}{c(\alpha,x)}\frac{dE}{dx}=\frac{\rho}{\epsilon},\label{eqn:Gauss Law}\\
E=-\nabla_{\alpha}V= -\frac{1}{c(\alpha,x)}\frac{dV}{dx}, \label{eqn:E del V}
\end{eqnarray}
where $c(\alpha,x)=\frac{\pi^{\alpha/2}}{\Gamma(\alpha/2)}|x|^{\alpha-1}$~\cite{zubair2017fractional}, $\rho$ is the carrier charge density, $\nu$ is the drift
velocity, $E$ is the electric field, and $\epsilon$ is the dielectric permittivity of the material, and $V$ is the electric potential.
Using $\nu=-\mu E$, Eq. (\ref{eqn:J=rho nu}) gives
\begin{eqnarray}
J=\rho \mu E \label{eqn:J=mu E}.
\end{eqnarray}

It should be emphasized that $\mu$ is an averaged quantity independent of space variables.

Now, inserting Eq. (\ref{eqn:Gauss Law}) into
Eq. (\ref{eqn:J=mu E}), we get
\begin{eqnarray}
J=\epsilon \mu \frac{1}{c(\alpha,x)}E\frac{dE}{dx},
\end{eqnarray}
which can be rewritten in the form of Bernoulli differential equation:
\begin{eqnarray}
\frac{dE}{dx}=\frac{c(\alpha,x)}{\epsilon \mu} J E^{-1}.
\label{eqn:Bernoulli}
\end{eqnarray}
Solving Eq. (\ref{eqn:Bernoulli}) with zero electric field condition (at SCL regime) at the injecting electrode, $E(0)=0$, we have
\begin{eqnarray}
E(x)= \sqrt{\frac{2Jx^{\alpha} \pi^{\alpha/2}}{\alpha \epsilon \mu
		\Gamma(\alpha/2)}}.
\label{eqn:BernoulliSolutionMGClassical}
\end{eqnarray}
By using Eq. (\ref{eqn:BernoulliSolutionMGClassical})  in solving Eq. (\ref{eqn:E del V}), we  obtain
\begin{eqnarray}
V(x)=
\sqrt{\left[\frac{\pi^{\alpha/2}}{\Gamma(\alpha/2)}\right]^3\frac{J}{\epsilon
		\mu }x^{3\alpha}}.
\label{eqn:VMGClassical}
\end{eqnarray}
which gives the modified MG law as a function of $\alpha$:
\begin{eqnarray}
J=
\frac{9\alpha^{3}}{8}\left[\frac{\Gamma(\alpha/2)}{\pi^{\alpha/2}}\right]^3\epsilon
\mu\frac{V_{0}^{2}}{L^{3\alpha}} .
\label{eqn:FDspaceMGClassical}
\end{eqnarray}
For $\alpha=1$, Eq. (\ref{eqn:FDspaceMGClassical}) reduces to the classical MG Law [Eq. (\ref{eqn:MGClassical})].

%

\subsection{Trap-limited (TL) model}
For a spatially disordered material with exponentially distributed traps in energy, the trap-limited TL-SCLC injection is derived here.
We assume that the mobility is field-independent, and that the density of the trapping states per unit energy range $h(E)$ above the valence band is described by the distribution
\begin{eqnarray}
h(E)=(N_{t}/kT_{c})\exp(-E/kT_{c}),
\label{eqn:h(E)}
\end{eqnarray}
where $E$ is the energy measured upward from the top of the valence band, $N_{t}$ is the total trap density, and $T_{c}$ is a characteristic constant of the distribution.
Following the Mark
and Helfrich (MH) approach~\cite{mark1962space}, we obtain
\begin{eqnarray}
\rho(x)= \left(\frac{J}{N_{c}\mu} \right) ^{1/l}e_{0}^{l-1/l}N_{t}E^{-1/l},
\label{eqn:h(E)}
\end{eqnarray}
where $l=T_{c}/T$ and $N_{c}$ is effective density of states. In this case the relation between free $n$ and trapped carrier density $n_{t}$ is given by
\begin{eqnarray}
\frac{n}{N}=\left(\frac{n_{t}}{N_{t}}\right)^l,
\label{eqn:n/N}
\end{eqnarray}
where, $N$ is total density of transport sites.
By solving Eq. (\ref{eqn:Gauss Law}), the governing equation is
\begin{eqnarray}
\frac{dE(x)}{dx}=F\frac{x^{\alpha-1}}{\left[ E(x)\right] ^{1/l}}\label{eqn:DE/dxtrap},\\
F=\left(\frac{J}{N_{c}\mu} \right) ^{1/l}e_{0}^{l-1/l}N_{t}/ \epsilon \frac{\pi^{\alpha/2}}{\Gamma(\alpha/2)}.
\end{eqnarray}
Integrating Eq. (\ref{eqn:DE/dxtrap}) on both sides gives
\begin{eqnarray}
E(x)= \left( \frac{l+1}{l}\frac{F}{\alpha}\right)^{\frac{l}{l+1}}x^{\frac{\alpha l}{l+1}},
\end{eqnarray}
and
\begin{eqnarray}
V(x)= \frac{\pi^{\alpha/2}}{\Gamma(\alpha/2)}\int E(x) x^{\alpha-1}.
\end{eqnarray}

The analytical evaluation of above integral leads to
\begin{eqnarray}
J&= &N_{c} \mu e^{1-l} \left[\frac{\Gamma(\alpha/2)}{\pi^{\alpha/2}}\right]^{2l+1} \left[ \frac{\epsilon \alpha l}{N_{t} (l+1)}\right]^{l} \nonumber\\
&&\times\left( \frac{2 \alpha l+ \alpha}{l+1} \right)^{l+1} \frac{V_{0}^{l+1}}{L^{2 \alpha l+\alpha}},
\label{eqn:Tl-SCLC-FD}
\end{eqnarray}
which reduces to Eq. (\ref{eqn:Tl-SCLC-Classical}) at $\alpha=1$.

It should be noted that for a Gaussian distribution of traps, a similar equation for the trap-limited current is derived, except that $l$ is then related to the depth and width of the trap distribution~\cite{hwang1976studies,abbaszadeh2016elimination}. In the case of Gaussian trap DOS centered at a distance $E_{tc}-E_{a}$ below the conduction-band edge ,the nondegenerate approximation gives~\cite{mandoc2007trap}
\begin{eqnarray}
\frac{n}{N}=\left(\frac{n_{t}}{N_{t}\exp((E_{tc}-E_{a})/kT_c)}\right)^l,
\label{eqn:n/N_guassian}
\end{eqnarray}
where, $E_{a}=\frac{\sigma^2}{2kT}$, $\sigma$ is the variance of Gaussian DOS. Finally, following the MH formalism, a current-voltage characteristic is obtained for Gaussian trap DOS, which is of the form
\begin{eqnarray}
J&= &N_{c} \mu e^{1-l} \nonumber \\
&&\times\left[\frac{\Gamma(\alpha/2)}{\pi^{\alpha/2}}\right]^{2l+1} \left[ \frac{\epsilon \alpha l}{N_{t} \exp((E_{tc}-E_{a})/kT_c )(l+1)}\right]^{l} \nonumber\\
&&\times\left( \frac{2 \alpha l+ \alpha}{l+1} \right)^{l+1} \frac{V_{0}^{l+1}}{L^{2 \alpha l+\alpha}},
\label{eqn:Tl-SCLC-FD-Gaussian}
\end{eqnarray}
which reduces to Eq. (7) in \cite{mandoc2007trap} at $\alpha=1$.

\subsection{Field-dependent mobility model}
If we simply combine the field-dependent mobility Eq. (\ref{eqn:field_dependence}), and Eq. (\ref{eqn:FDspaceMGClassical}), the modified SCL model of field-dependent mobility for spatially disordered semiconductors is

\begin{eqnarray}
J=
\frac{9\alpha^{3}}{8}\left[\frac{\Gamma(\alpha/2)}{\pi^{\alpha/2}}\right]^3\epsilon
\mu_{0}\frac{V_{0}^{2}}{L^{3\alpha}}\exp(0.89 \gamma \sqrt{E}).
\label{eqn:MGClassicalFD_fielddep}
\end{eqnarray}
where $\gamma$ is just a fitting parameter. In general, to include the field dependence of the mobility in SCLC model, coupled equations such as Eqs. (\ref{eqn:J=rho nu}-\ref{eqn:E del V}) must be solved consistently~\cite{chen1978model}.
It is however possible to derive an analytic solution if the field dependence of the drift mobility can be expressed in power law~\cite{abkowitz1993time} given by
\begin{eqnarray}
\mu=\mu_{0} \left( \frac{E}{E_{0}}\right) ^{n},
\label{eqn:field_dependence_powerlaw}
\end{eqnarray}
with $\mu=\mu_{0} $ at $E=E_{0}$.
By using this power law of mobility, we solve Eqs. (\ref{eqn:J=rho nu}-\ref{eqn:E del V}) to obtain an analytical solution of
\begin{eqnarray}
J&=&
\frac{\epsilon
	\mu_{0}}{E_{0}^{n}}\left[\frac{\Gamma(\alpha/2)}{\pi^{\alpha/2}}\right]^{n+3}
\left[\frac{n\alpha-n+\alpha}{n+2}\right]\nonumber\\&&\times
\left[\frac{2n\alpha+3\alpha-n}{n+2}\right]^{n+2}
\frac{V_{0}^{n+2}}{L^{2n\alpha+3\alpha-n}},
\label{eqn:PowerLawFieldMobility-SCLC-FD}
\end{eqnarray}
which reduces to Eq. (\ref{eqn:FDspaceMGClassical}) at $n=0$.

\section{\label{sec: Results} Results and Discussions}
By analyzing the thickness ($L$) dependence of the classical SCLC models, we see the dependence of $L^{-3}$ (at fixed $V$),  $L^{-2l-1}$ (at fixed $V$) and $L^{-1}$ (at fixed $E$), respectively, from Eq. (\ref{eqn:MGClassical}), Eq. (\ref{eqn:Tl-SCLC-Classical}), and Eq. (\ref{eqn:MGClassical_fielddep}).
However, as predicted by corresponding modified SCLC  models in Sec. II, the thickness dependence will be reduced by the fractional-dimension parameter $\alpha$, which accounts to the spatial disorder in the underlying solids. In other words, the thickness dependence of the modified SCLC models will provide a tool to characterize the spatial disorder in the porous organic semiconductor.

\begin{table*}[!ht]
		\caption{Thickness dependence of organic semiconductor devices based on disordered semiconducting organic polymers and its relation to parameter $\alpha$ used in modified SCLC models.} 
	\begin{center}
		\label{tab:data11}
		\begin{tabular}{lllll}
			\hline
			\hline
			Polymer Type                    & \begin{tabular}[c]{@{}l@{}}Average thickness \\ dependence \\ (extracted from \\ experiment)\end{tabular} & \begin{tabular}[c]{@{}l@{}}Thickness dependence \\ relation\\ (modified MG law)\end{tabular} & \begin{tabular}[c]{@{}l@{}}Calculated $\alpha$ \\ (measure of spatial \\ disorder)\end{tabular} & \begin{tabular}[c]{@{}l@{}}Ref. \\ (experimental \\ data)\end{tabular} \\

			\hline
			\hline
			NRS-PPV                         & 2.895 (Fig. (3a))                                                                                               & 3$\alpha$  (Eq. (19))                                                                                & 0.965                                                       & \cite{blom2005thickness}                 \\
			PFO                             & 2.9 (Fig. (3b))                                                                                                    & 3$\alpha$                                                                                 & 0.967                                                       & \cite{nicolai2010space}                 \\
			$\alpha$-conjugated-Sexithienyl & 2.6 (Fig. (3c))                                                                                                    & 3$\alpha$                                                                                 & 0.86                                                        & \cite{horowitz1990evidence}                 \\
			poly-fluorene-based             & 2.5 (Fig. (3d))                                                                                                    & 3$\alpha$                                                                                 & 0.83                                                        & \cite{coehoorn2006measurement}                \\
			$OC_{1}C_{10}-PPV$              & 11 (Fig. (4a))                                                                                                     & 2$\alpha l$+$\alpha$ (Eq. (27))                                                                         & 0.918                                                       & \cite{mandoc2006electron}                 \\
			NPB                             & 0.52 (Fig. (4b))                                                                                                   & 3$\alpha$-2 (Eq. (30))                                                                                 & 0.84                                                        & \cite{chu2007hole}              \\
			\hline
			\hline
		\end{tabular}
	\end{center}
\end{table*}
\subsection{Implications of modified SCLC model on mobility extraction}
\begin{figure}[!ht]
	\centering
	\includegraphics[width=.3\textwidth]{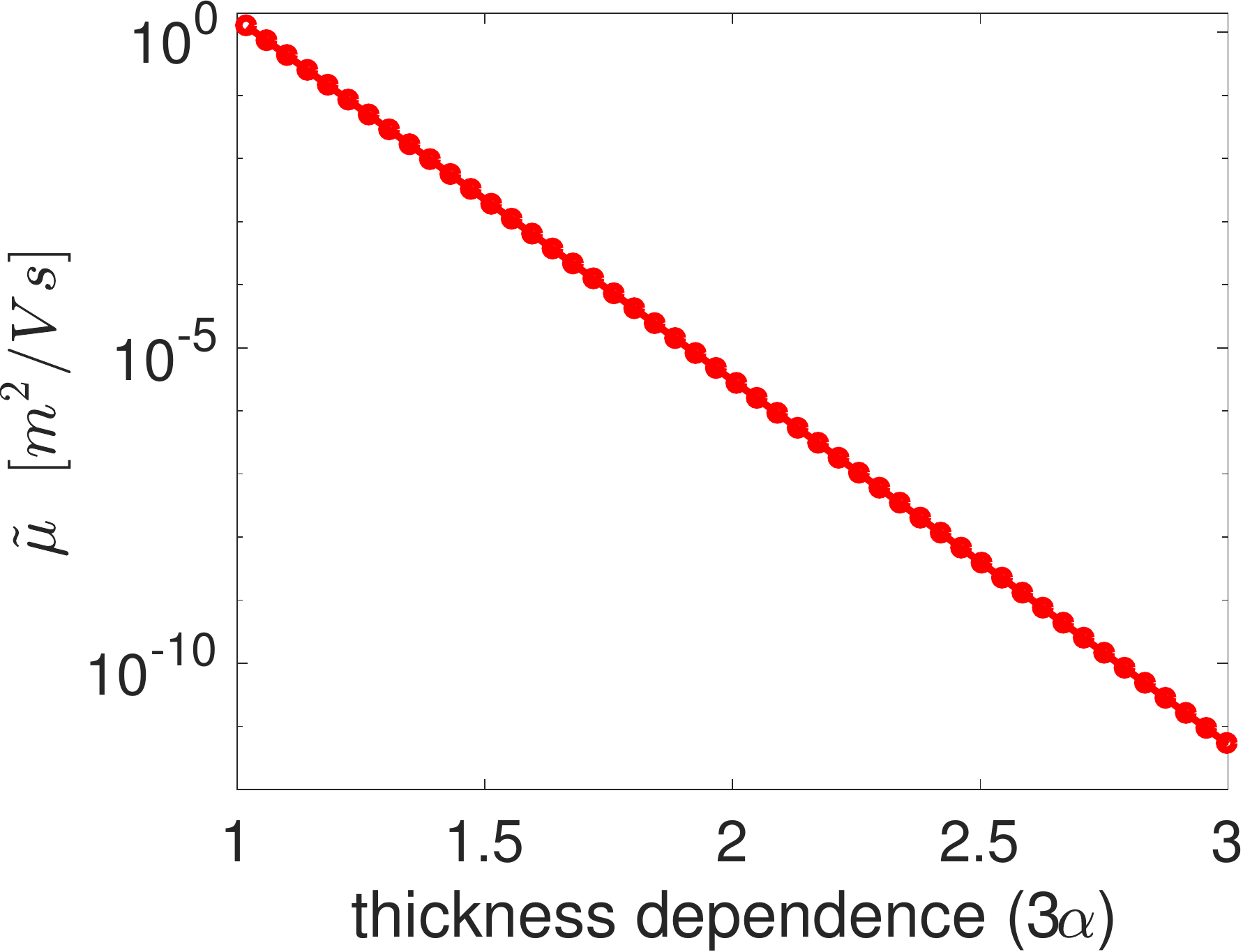}
	\caption{The variation of the mobilities of PPV based devices as a function of varying thickness dependence ($L^{3 \alpha}$). It is shown that a variation in thickness dependence leads to several orders of under- or over-estimation of mobility values. The J-V data is taken from~\cite{blom2005thickness} for $L=950 $ nm.}
	\label{fig:extract_mobility}
\end{figure}

Before proceeding with the analysis of thickness dependence in some
reported experimental data, it would be of interest to see the
effect of variation in thickness dependence due to spatial disorder
in the semiconductors on the mobility values extracted form
experimental J-V curves taken from~\cite{blom2005thickness}. We denote the extracted values of mobility by $\tilde{\mu}$, to distinguish them from actual mobility $\mu$ for this device. In Fig.
(\ref{fig:extract_mobility}), the extracted mobility ($\tilde{\mu}$) is plotted as a function of thickness dependence parameter $3\alpha$.
The fractional-dimension parameter $\alpha$ corresponds to the
measure of spatial disorder in the semiconductor, with $\alpha=1$
corresponding to zero spatial disorder. It can be seen that $\tilde{\mu}$ is sensitively influenced by $\alpha$. Thus it is important to check the $L$ dependence rather than assuming $\alpha = 1$ which may no longer be valid for complicated materials such as
	porous and amorphous organic materials.

\subsection{Consistency of $\alpha $ extracted from experimental data}
\begin{figure}[!hb]
	\subfloat[\label{fig:enhancementxo}]{%
		\includegraphics[width=.22\textwidth]{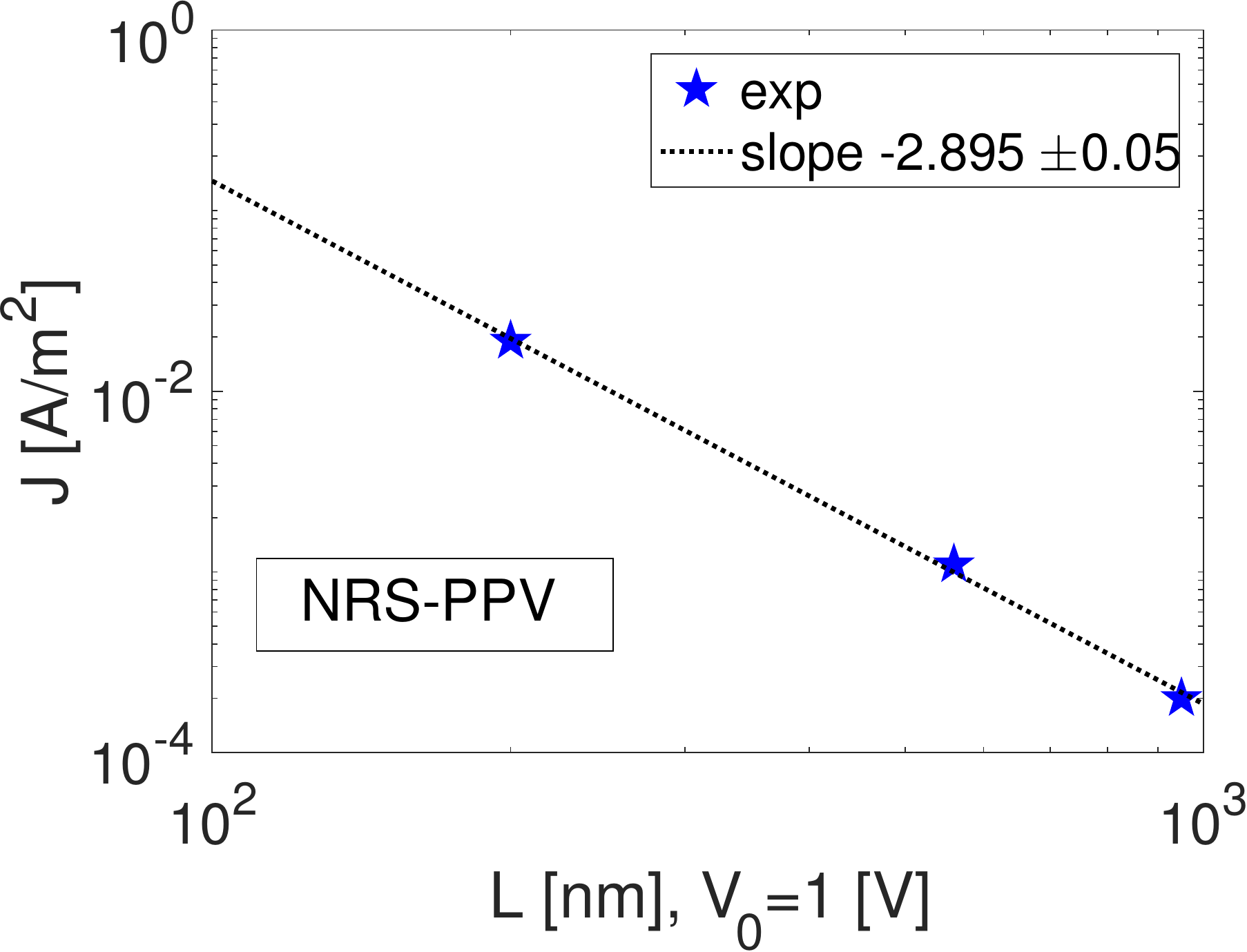}%
		\label{Fig:Slopes1}
	}
	\subfloat[\label{fig:enhancementalpha}]{%
		\includegraphics[width=.22\textwidth]{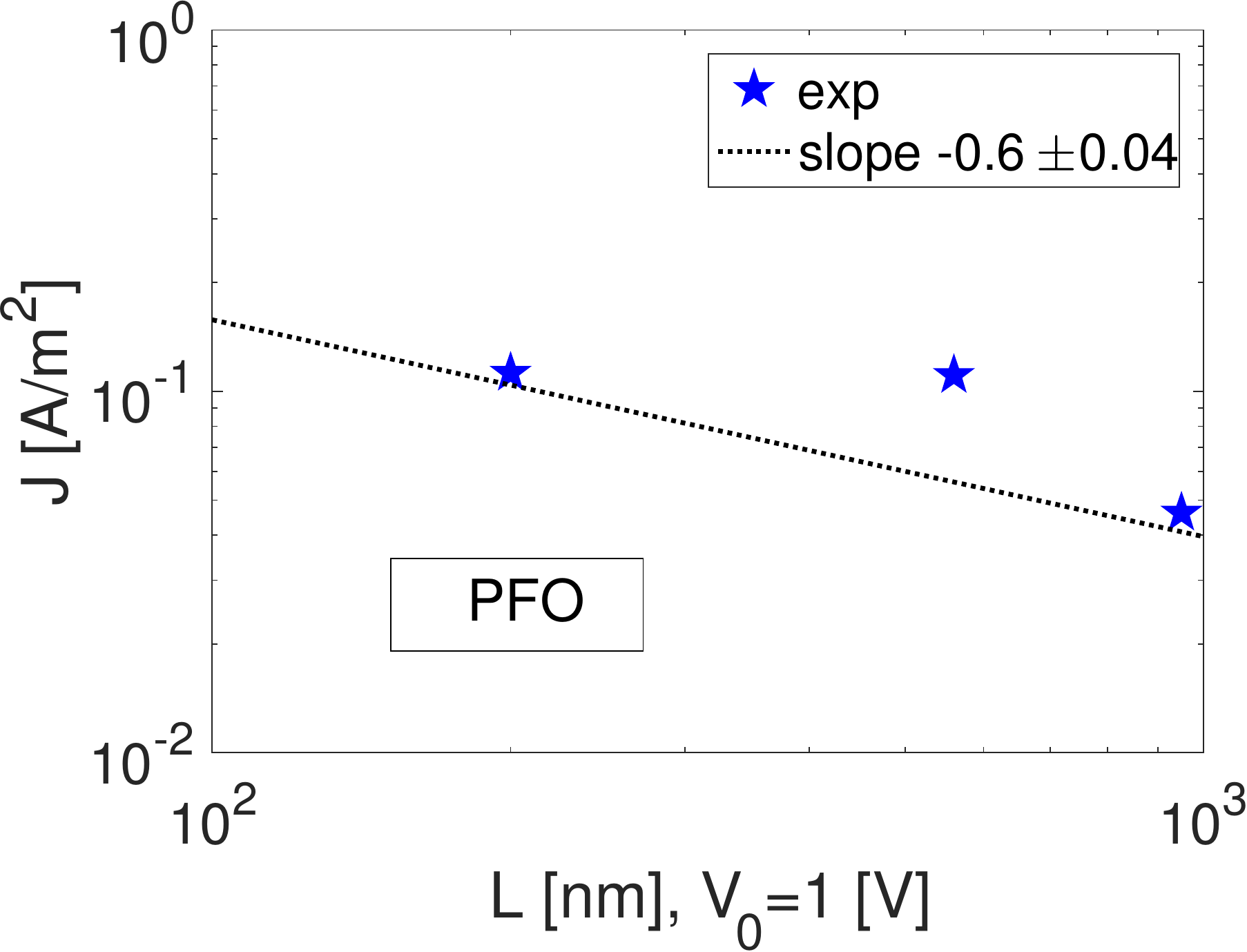}%
		\label{Fig:Slopes2}
	} 
	\hfill
	\subfloat[\label{fig:enhancementxo}]{%
		\includegraphics[width=.22\textwidth]{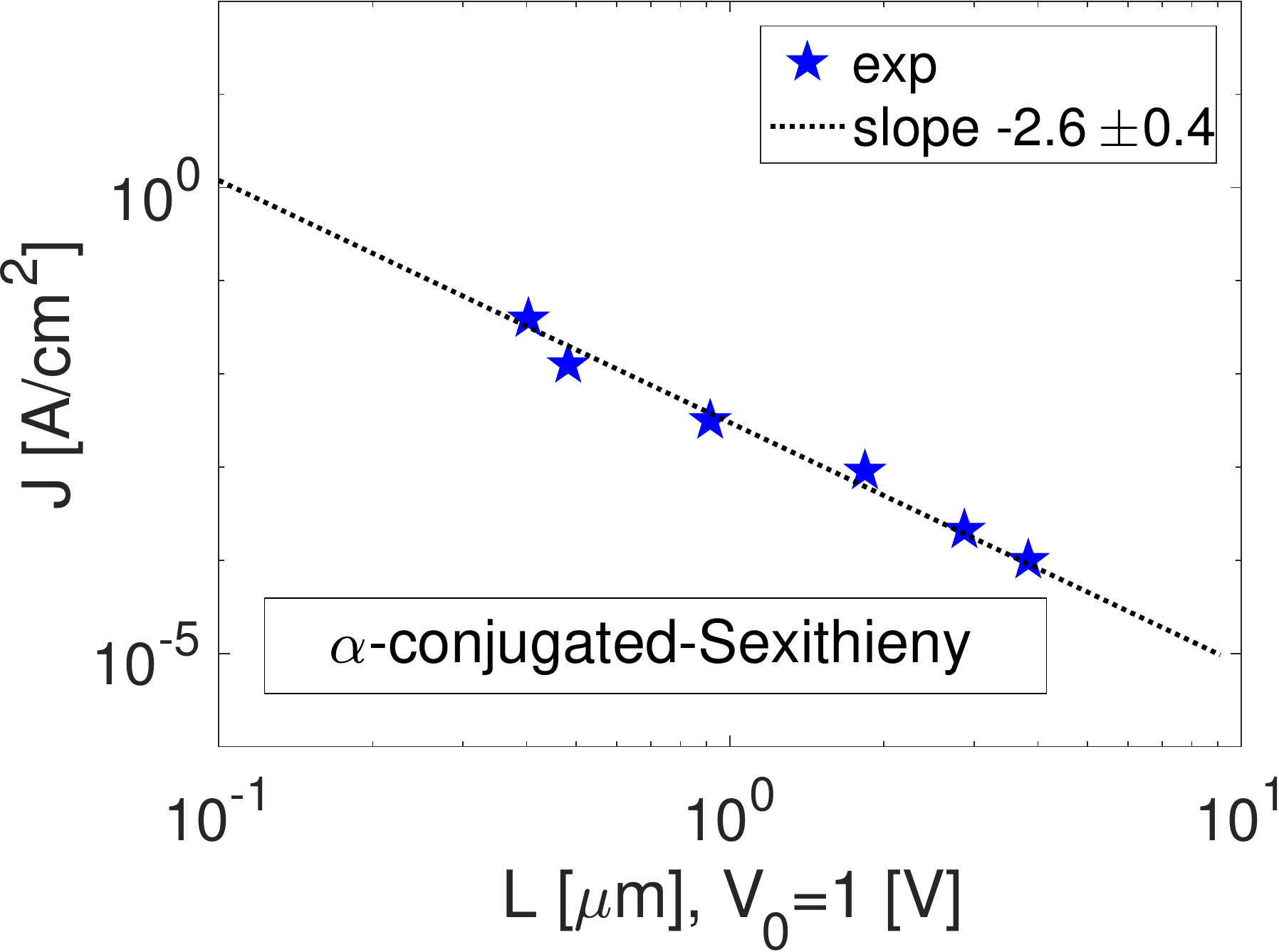}%
		\label{Fig:Slopes3}
	}
	\subfloat[\label{fig:enhancementalpha}]{%
		\includegraphics[width=.22\textwidth]{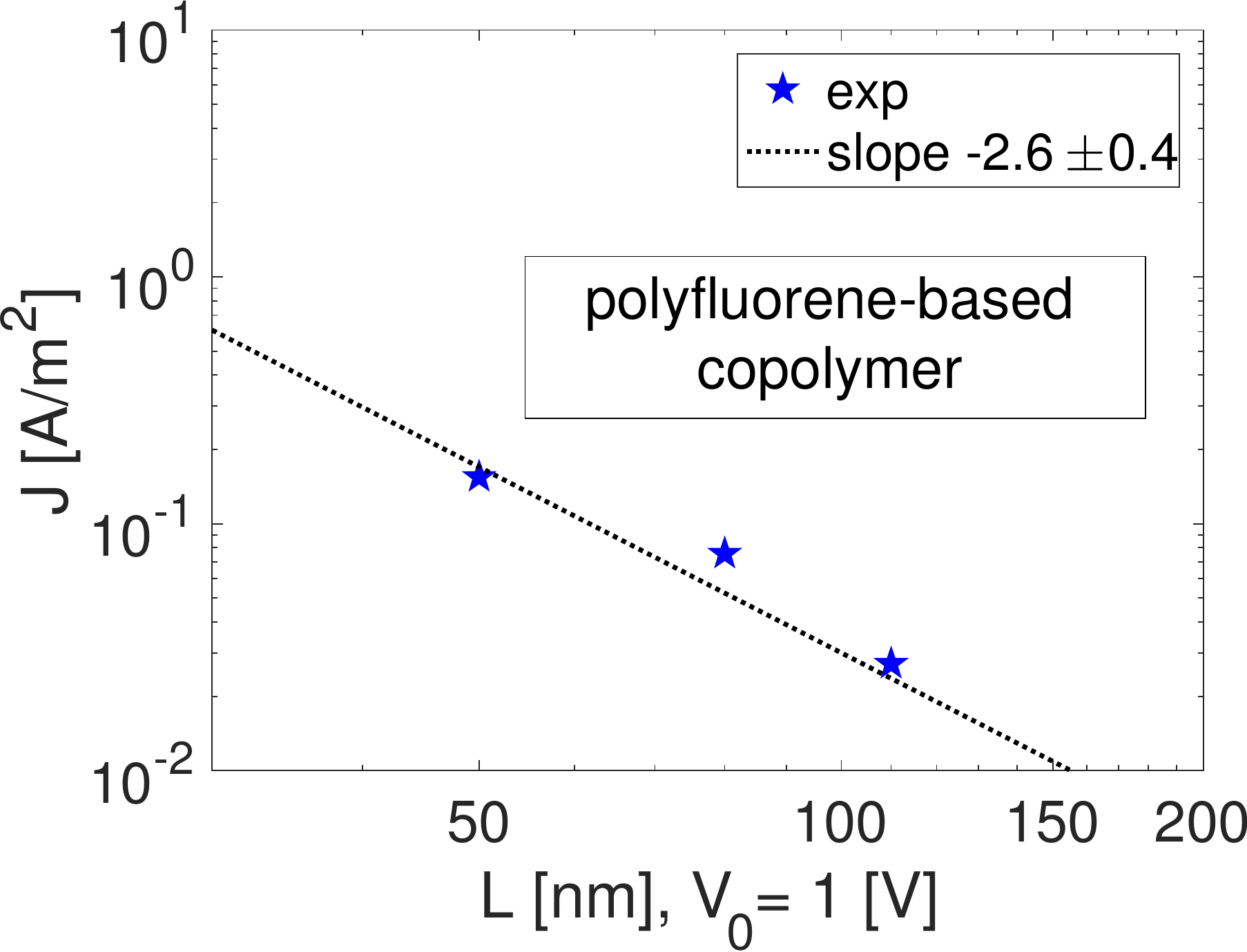}%
		\label{Fig:Slopes4}
	}\caption{The thickness dependency of the current density at fixed voltage for different polymers taken from current-voltage characteristics in the literature: \textbf{(a)} $NRS-PPV$~\cite{blom2005thickness} \textbf{(b)}
	$PFO$~\cite{nicolai2010space} \textbf{(c)}
	$\alpha$-conjugated-Sexithienyl~\cite{horowitz1990evidence} \textbf{(d)}
	poly-fluorene-based~\cite{coehoorn2006measurement}.} \label{Fig:Slopes}
\end{figure}

Most organic semiconductors have spatial as well as energetic disorder. The existing mobility models incorporate combined effect of energetic and spatial disorder using a range of mobility models including field- and density-dependent mobility. However, our model predicts that the spatial disorder can affect the thickness scaling of SCLC. Here, we explore the available experimental data of SCLC versus device thickness for a range of disordered organic semiconductors. It is observed that the thickness scaling of SCLC varies as predicted by our model, to which not much attention was paid previously and it was assumed trivially that thickness scaling follows standard MG law which turns out to be not true for organic semiconductor in several example cases reported in the following. Fig. (\ref{Fig:Slopes}) shows the corresponding thickness $L$
dependence for various devices using different organic materials.
The results shown in Fig. (\ref{Fig:Slopes1}-\ref{Fig:Slopes4}) are
at fixed voltage regime (constant mobility) with varying thickness
$L$. Based on the classical MG law, we will expect a scaling of
$L^{-3}$. However, due to spatial disorder, the results show a
weaker thickness dependence in the range $L^{-2}$ to $L^{-3}$, which
corresponds to about $\alpha\approx0.8$ to $\alpha\approx1$.

\begin{figure}[!hb]
	\subfloat[\label{fig:enhancementxo}]{%
		\includegraphics[width=.22\textwidth]{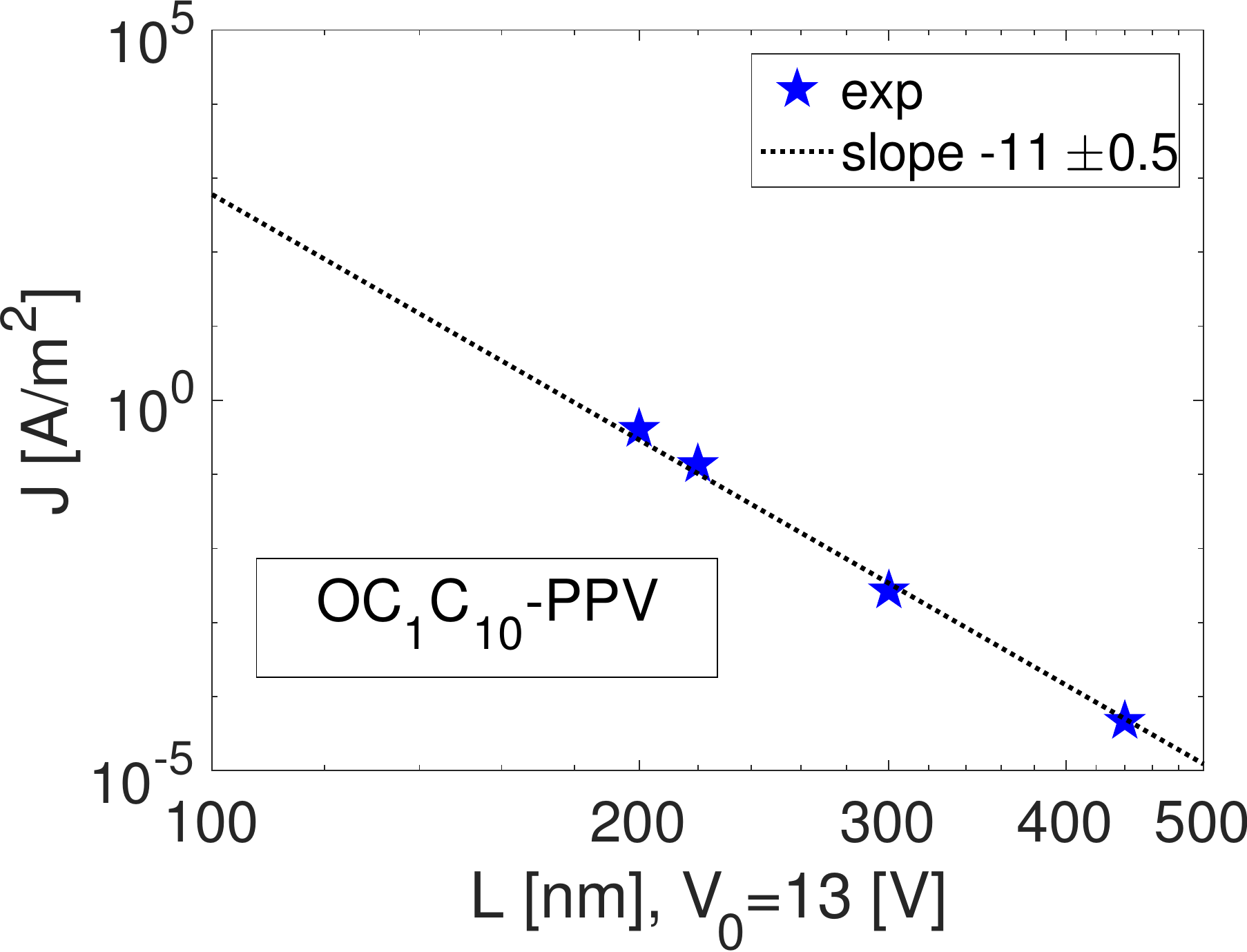}%
		\label{Fig:Slopes5} }
	\subfloat[\label{fig:enhancementalpha}]{%
		\includegraphics[width=.22\textwidth]{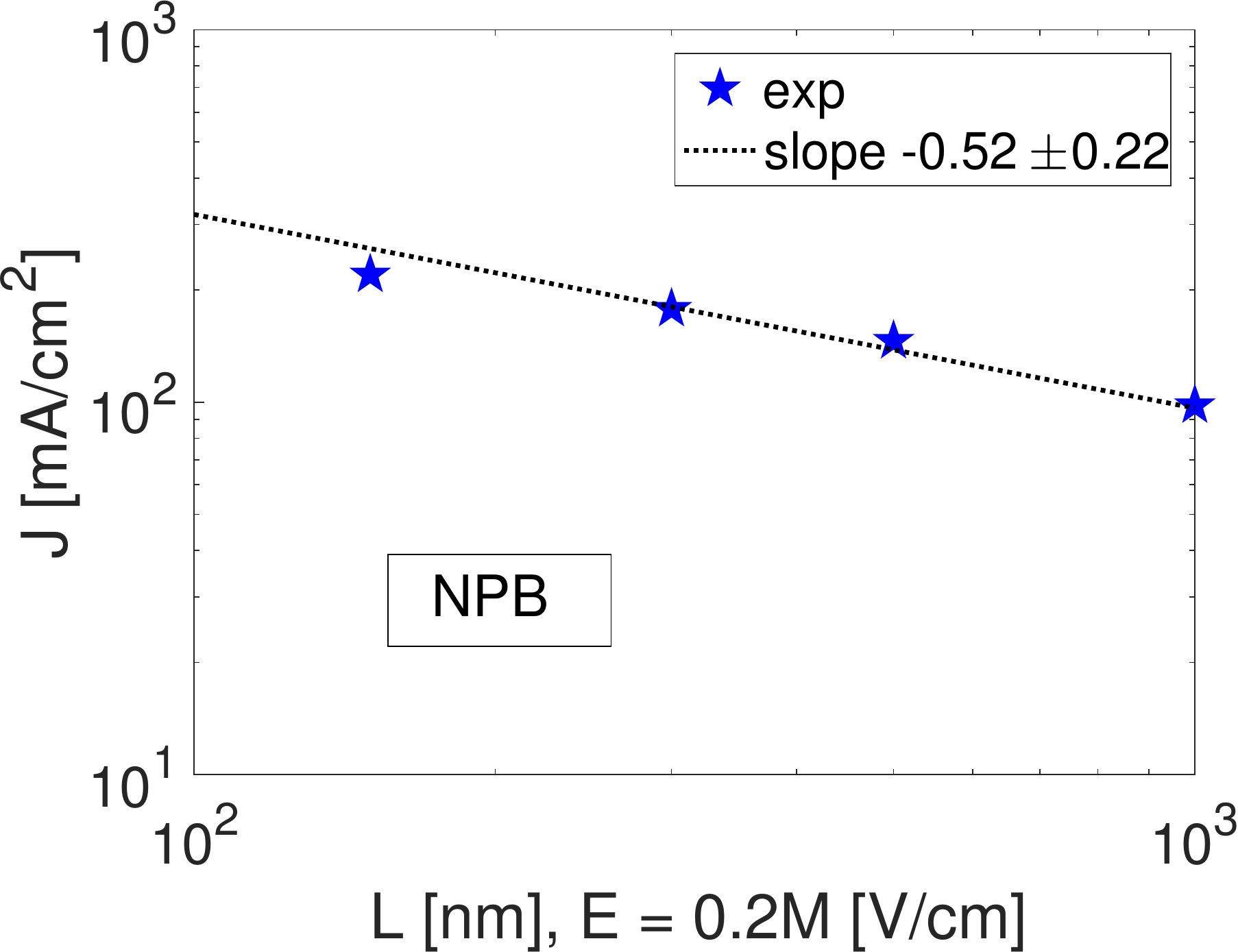}%
		\label{Fig:Slopes6}
	}\caption{The thickness dependency of \textbf{(a)} trap-limited current density at fixed voltage for polymer $OC_{1}C_{10}-PPV$ taken from current-voltage characteristics in the literature~\cite{mandoc2006electron} \textbf{(b)} current density at fixed electric field for polymer $NPB$ taken from current-field characteristics in the literature~\cite{chu2007hole}.} \label{Fig:Slopes_remaining}
\end{figure}

For the results shown in Fig. (\ref{Fig:Slopes5}) for a trap-filled
organic material we have $l=T_{t}/T=1500/273=5.49$. Based on the
classical TL-SCLC model (without any spatial disorder), the thickness
dependence should be $L^{-11.98}$. However, the experimental fitting
shows again a weaker dependence, which corresponds to $\alpha=0.918$
based on Eq. (\ref{eqn:Tl-SCLC-FD}), instead of $\alpha$ = 1. For results shown in Fig. (\ref{Fig:Slopes6}), the $L$ scaling is calculated from measurements at fixed electric field. As mentioned earlier for
field-dependent mobility, the thickness dependence should be
$L^{-1}$ at a fixed field for negligible spatial disorder at
$\alpha$ = 1. However, we observe a weaker dependence, which
corresponds to $\alpha=0.86$ based on
Eq.(\ref{eqn:MGClassicalFD_fielddep}). Table I summarizes the
results of Fig. (\ref{Fig:Slopes}-\ref{Fig:Slopes_remaining}) for various disordered organic
semiconductor based devices and its relation to the $\alpha$
parameter used in our models to fit with the experimental results.
Our analysis suggests that the traditional $L$ scaling formulated in
the classical SCLC models may not be suitable for organic
semiconductors, and it will provide an inaccurate estimation of
the mobility if such models are used. Note that the uncertainty in the
measurement of $L$, which is about 5 nm from normal experimental
setup, is not able to explain the variation from the expected
$\alpha$ = 1 assuming the classical models are correct.

\begin{figure}[!hb]
	\subfloat[\label{fig:enhancementxo}]{%
		\includegraphics[width=.22\textwidth]{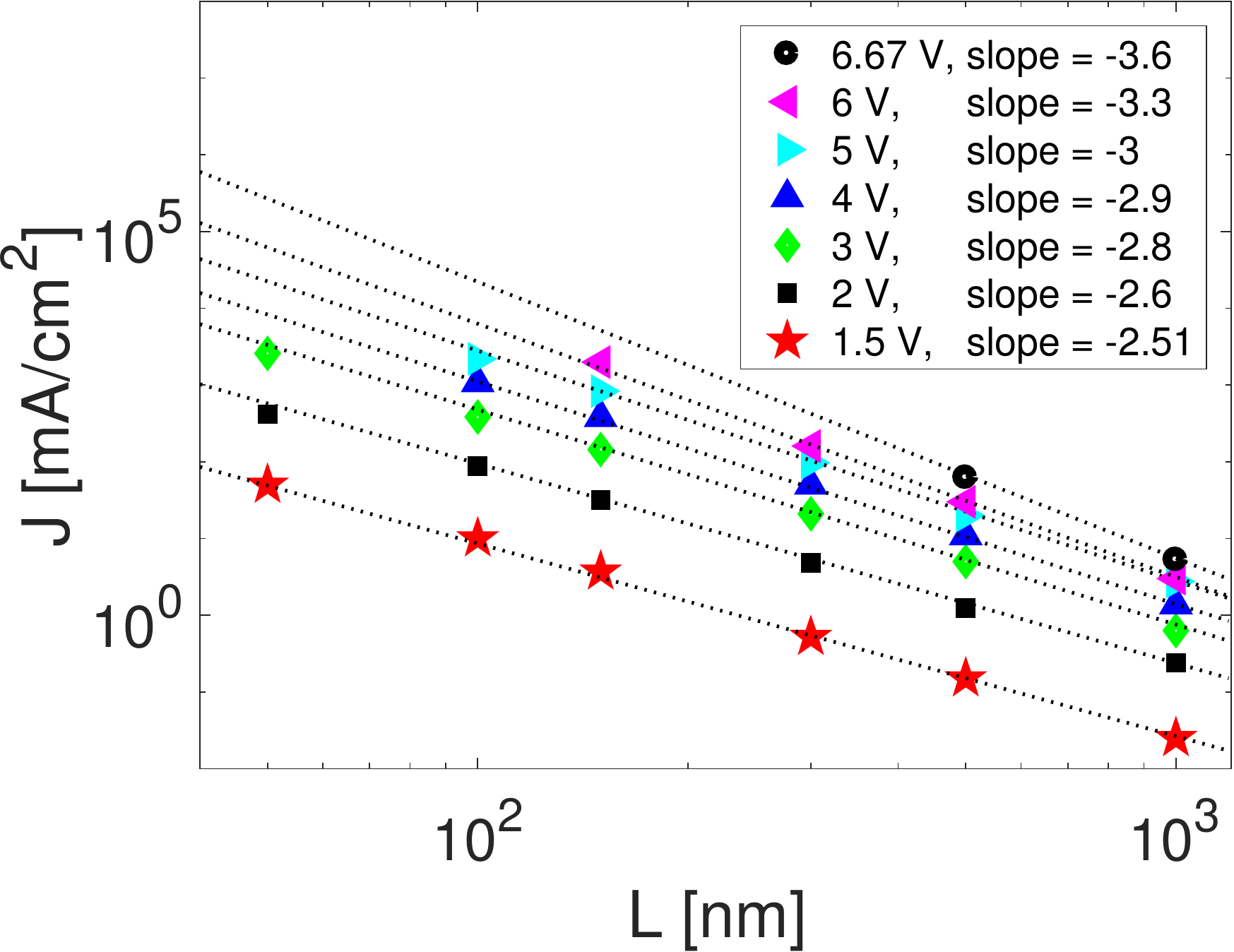}%
		\label{Fig:JvsL_fixedV}
	}
	\hfill
	\subfloat[\label{fig:enhancementalpha}]{%
		\includegraphics[width=.22\textwidth]{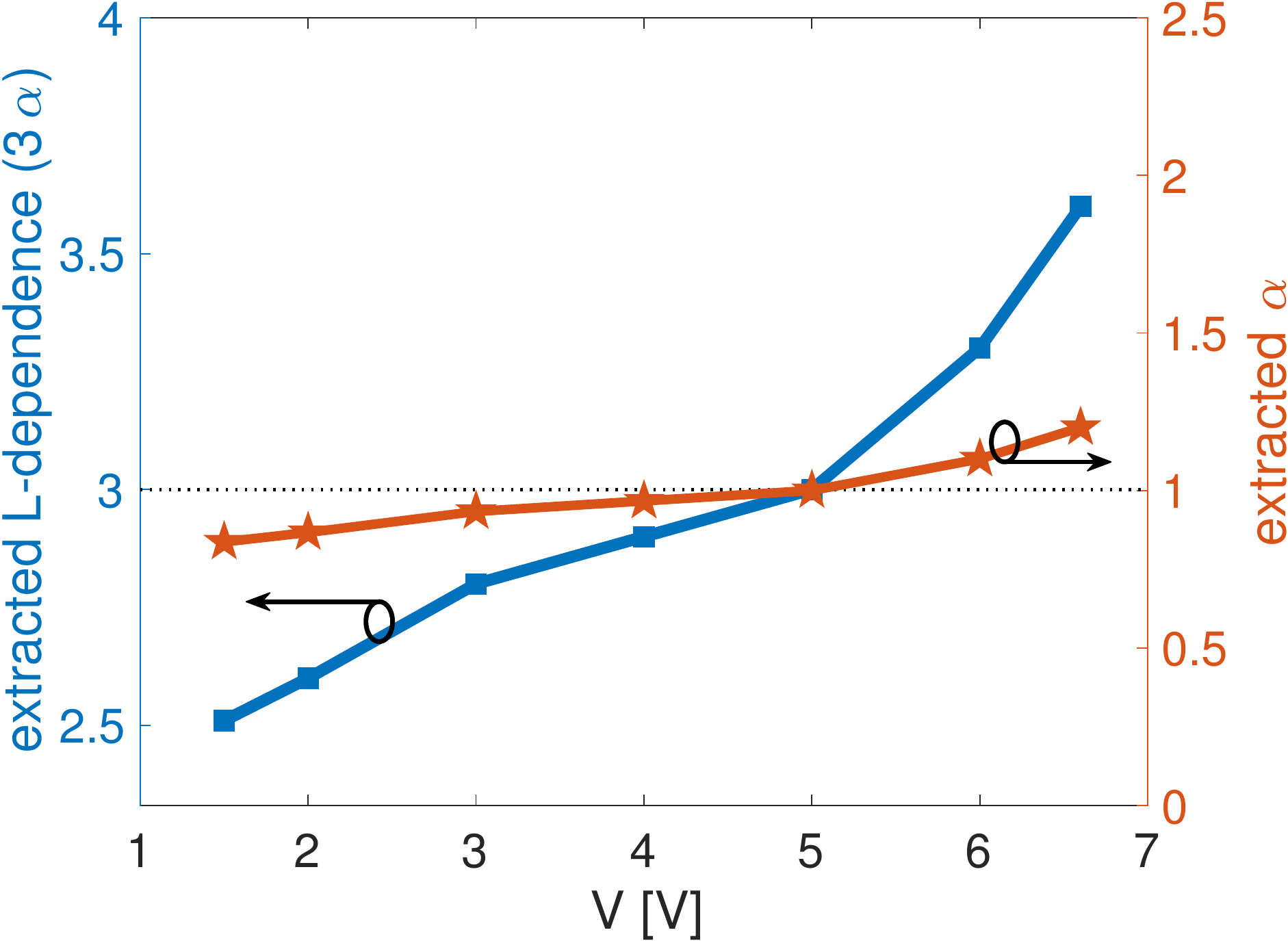}%
		\label{Fig:Alpha_fixedV}
	} 
	\hfill
	\subfloat[\label{fig:enhancementxo}]{%
		\includegraphics[width=.22\textwidth]{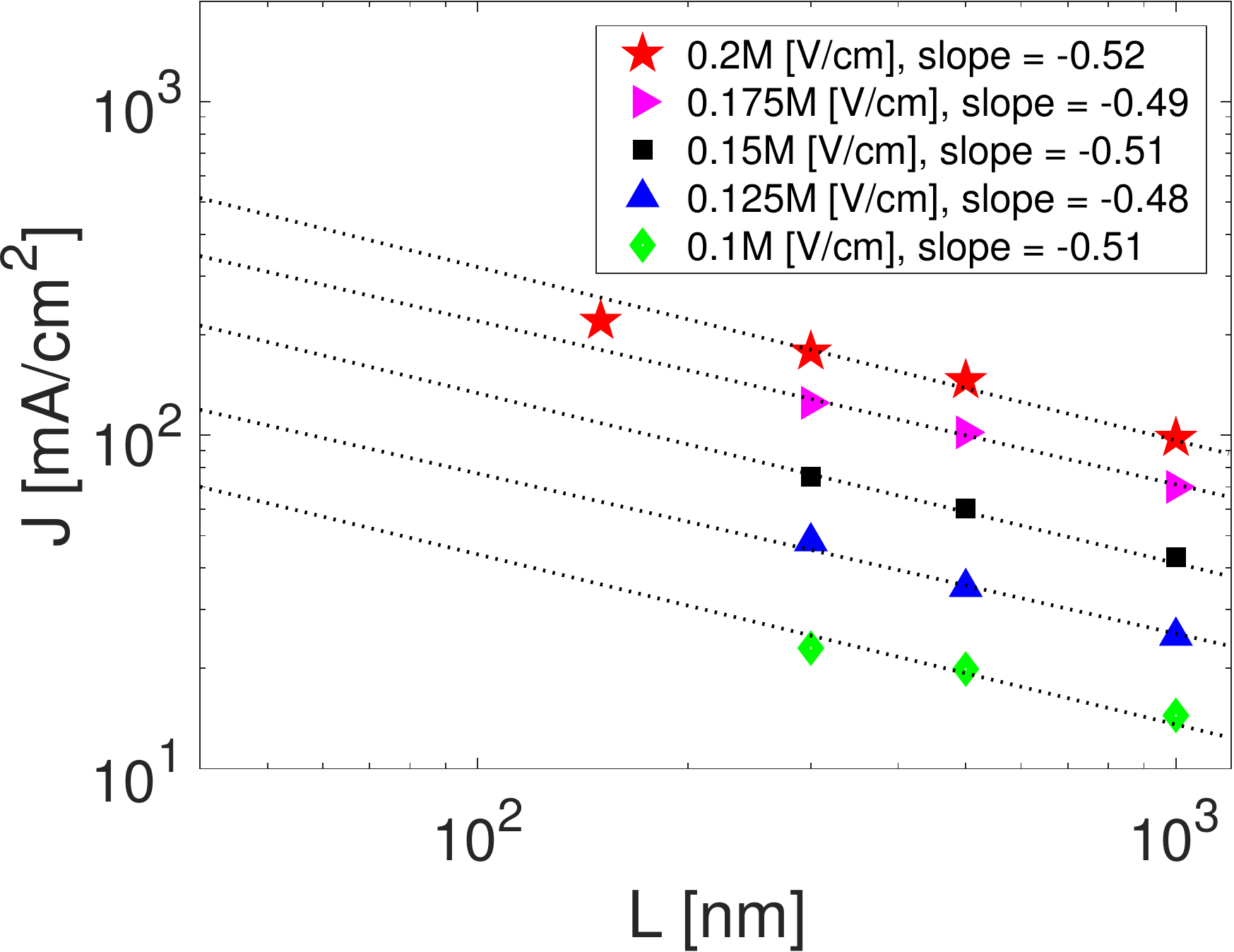}%
		\label{Fig:JvsL_fixedE}
	}
	\hfill
	\subfloat[\label{fig:enhancementalpha}]{%
		\includegraphics[width=.22\textwidth]{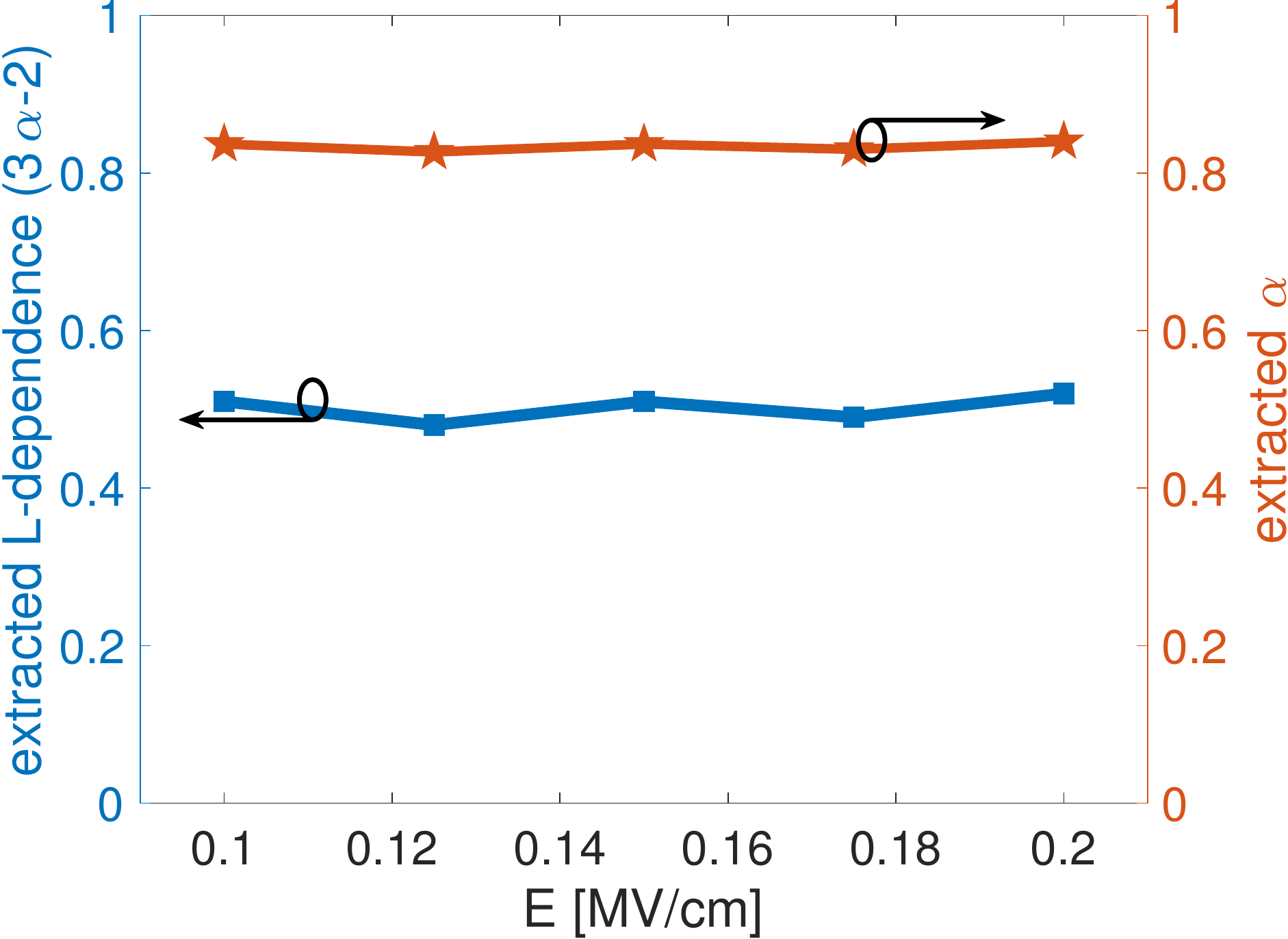}%
		\label{Fig:Alpha_fixedE}
	}\caption{Experimental data for polymer $NPB$ based devices taken from~\cite{chu2007hole}.  \textbf{(a)} The current density versus device thickness for varying voltages (applied voltage $V$ and slope is shown in legend). \textbf{(b)} The extracted thickness dependence and the parameter $\alpha$ at different voltages. \textbf{(c)} The current density versus device thickness for varying electric field (applied electric field $E$ and slope is shown in legend). \textbf{(d)} The extracted thickness dependence and the fractional dimension parameter $\alpha$ at different applied electric fields.} \label{Fig:Extraction_of_alpha}
\end{figure}

In Fig. (\ref{Fig:Slopes}), we have extracted the  thickness scaling of SCLC at low voltages to avoid the field-dependent SCLC regime. However, one must be careful while extracting the slopes at higher applied voltages. As the SCLC is \textit{field-dependent} at high applied voltage, the extraction of $\alpha$ should be performed at fixed electric field strength, $E$, rather than at fixed voltage, $V$. To demonstrate this, we analyze the SCLC versus voltage data of polymer NPB based devices reported in Ref.~\cite{chu2007hole}, and plot the current density against device thickness at different voltages in Fig. (\ref{Fig:JvsL_fixedV}) . The varying slope of $J$ versus $L$ shows that the thickness dependence of SCLC varies at different applied voltages due to $\exp(\gamma \sqrt{V/L})$ factor in field-dependent mobility model of Eq. (\ref{eqn:MGClassicalFD_fielddep}). Fig. (\ref{Fig:Alpha_fixedV}) shows the extracted thickness dependence at different voltages which immediately reveals that the value of the extracted $\alpha$ is inconsistent at different voltage. At high-voltage regime where field-dependence becomes non-negligible, the extracted thickness dependence even becomes stronger than $L^{-3}$ which leads to an unphysical value of $\alpha>1$. This clearly reveals the fallacy of extracting $\alpha$ from the J-L curve at fixed voltage. Instead, the $\alpha$ should be extracted at fixed \textit{electric field strength} as indicated by Eq. (\ref{eqn:MGClassicalFD_fielddep}), i.e., $J \propto L^{3\alpha-2}$ at fixed $E$. Fig. (\ref{Fig:JvsL_fixedE}) and (\ref{Fig:Alpha_fixedE}) shows the J-L characteristics and the extracted $\alpha$ at different $E$, respectively. In this case, a singular value of $\alpha\approx 0.84$ is extracted for all applied electric field strengths. More importantly, this value of $\alpha$ is consistent with that extracted from the low-voltage regime of Fig. (\ref{Fig:Alpha_fixedV}).

\begin{figure}[!hb]
	\subfloat[\label{fig:APL}]{%
		\includegraphics[width=.22\textwidth]{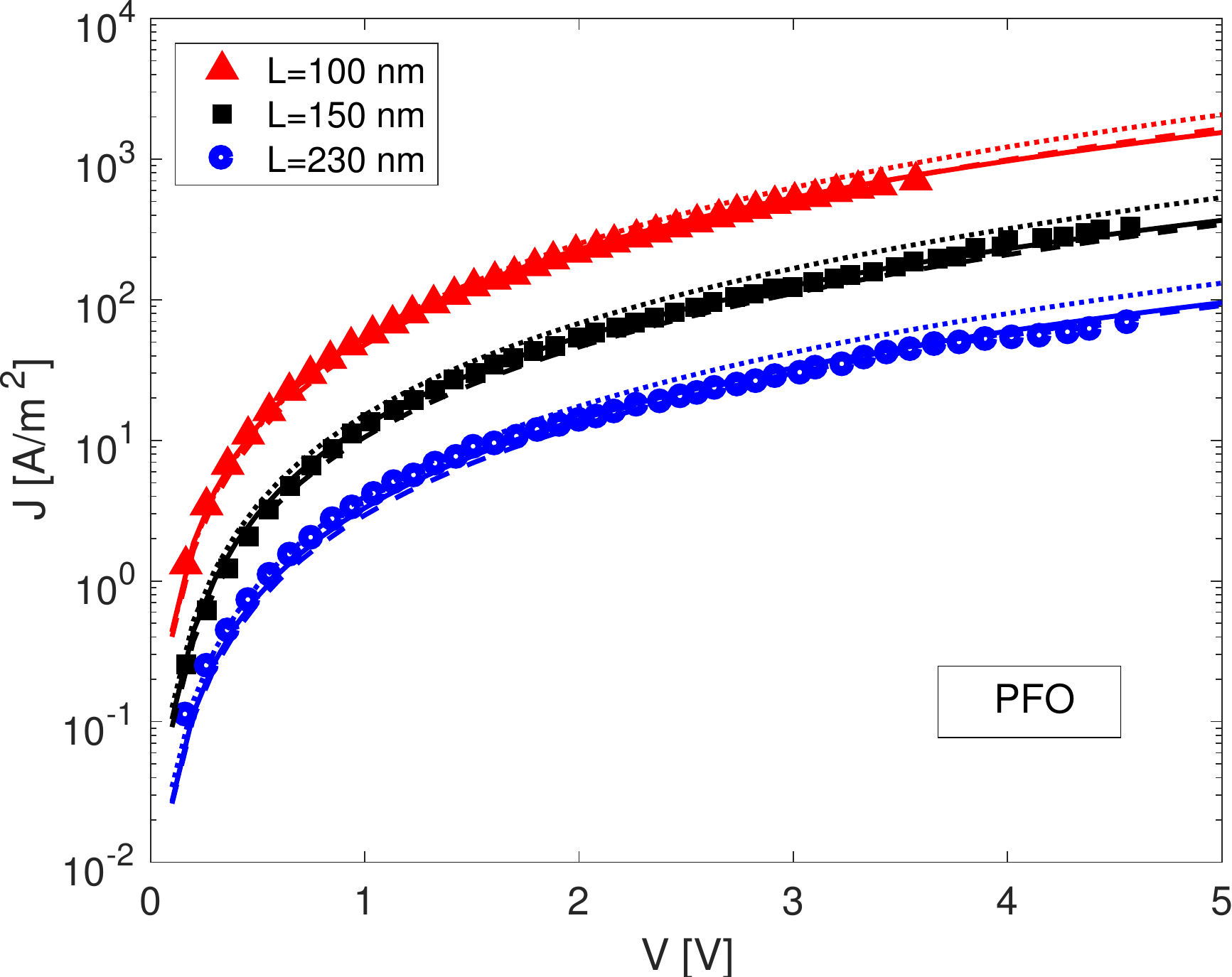}%
		\label{fig:APL96_2010} }
	\subfloat[\label{Fig:APL}]{%
		\includegraphics[width=.22\textwidth]{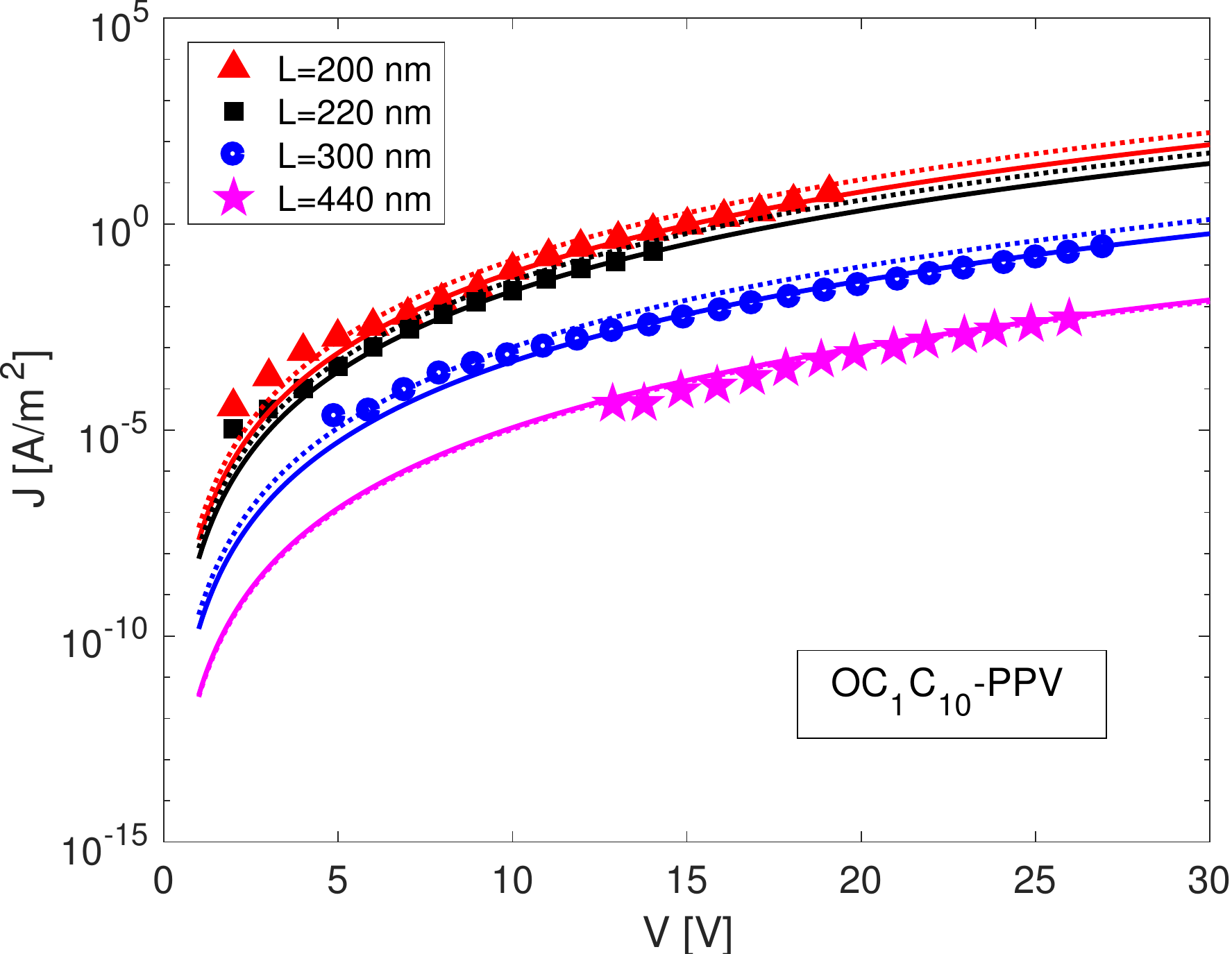}%
		\label{Fig:PRB73_2010}
	}\caption{\textbf{(a)} Room-temperature current density vs voltage characteristics data from~\cite{nicolai2010space} for $PFO$ diodes with thicknesses of 100 (red), 150 (black), and 250 (blue) nm, respectively. Experiment (circles), Eq. \ref{eqn:MGClassical_fielddep} (dotted lines), Eq. \ref{eqn:MGClassicalFD_fielddep} (dashed lines), Eq. \ref{eqn:PowerLawFieldMobility-SCLC-FD} (solid lines). The parameters used are $\gamma=1.2\times10^{-4}(m/V)^{1/2}$, $\mu_{0}=1.3\times10^{-9} m^2/Vs$,  $\alpha=0.967$, $E_{0}=0.1/L^{\alpha}$, $n=0.088$. \textbf{(b)} Room-temperature current density vs voltage characteristics data from~\cite{mandoc2006electron} for $OC_{1}C_{10}-PPV$ diodes with thicknesses of 200 (red), 220 (black), 300 (blue), and 440 (green) nm respectively. Experiment (circles), Eq. \ref{eqn:Tl-SCLC-Classical} (dotted lines), Eq. \ref{eqn:Tl-SCLC-FD} (solid lines). The parameters used are $\alpha=0.918$, $N_{t} = 8.5 \times10^{23} m^{-3}$, $T_{t} = 1500 K$ and zero-field mobility of $5\times10^{-11} m^2/Vs$.} 
\end{figure}

In Fig. (\ref{fig:APL96_2010}), the room-temperature current density versus voltage characteristics data from Ref.~\cite{nicolai2010space} is shown for PFO diodes of varying thickness together with various numerical models calculations. It should be noted that the classical model of Eq. (\ref{eqn:MGClassical_fielddep}) requires different values of $\gamma$ to be used in order to fit with the experimental data despite the fact that the devices are composed of the same type of polymer. To address this inconsistency, we fitted the experimental data using our modified SCLC model with $\alpha = 0.967$ [extracted from Fig. (\ref{Fig:Slopes2})]. Remarkably, our modified model is able to fit the experimental J-V curves of all devices using a singular consistent value of $\gamma = 1.2\times 10^{-4} (m/V)^{1/2}$. Similarly, in Fig. (\ref{Fig:PRB73_2010}) the room-temperature current density versus voltage data from Ref.~\cite{mandoc2006electron} for $OC_{1}C_{10}-PPV$ diodes with varying thicknesses is shown. The classical model in Eq. (\ref{eqn:Tl-SCLC-Classical}) fails to reproduce the experimental results with $N_{t}$ fixed for all L. In contrast, by using our modified trap-limited SLC model with $\alpha = 0.918$ extracted from Fig. (\ref{Fig:Slopes5}), a much better agreement with experimental results is obtained at a fixed $N_{t}$. These results show that the modified MG law can sufficiently describe the thickness dependence of SCLC for given range of applied voltages.



\subsection{Fitting experimental current-voltage characteristics and mobility extraction using modified SCLC model }
\begin{figure}[!hb]
	\subfloat[\label{fig:APL}]{%
		\includegraphics[width=.22\textwidth]{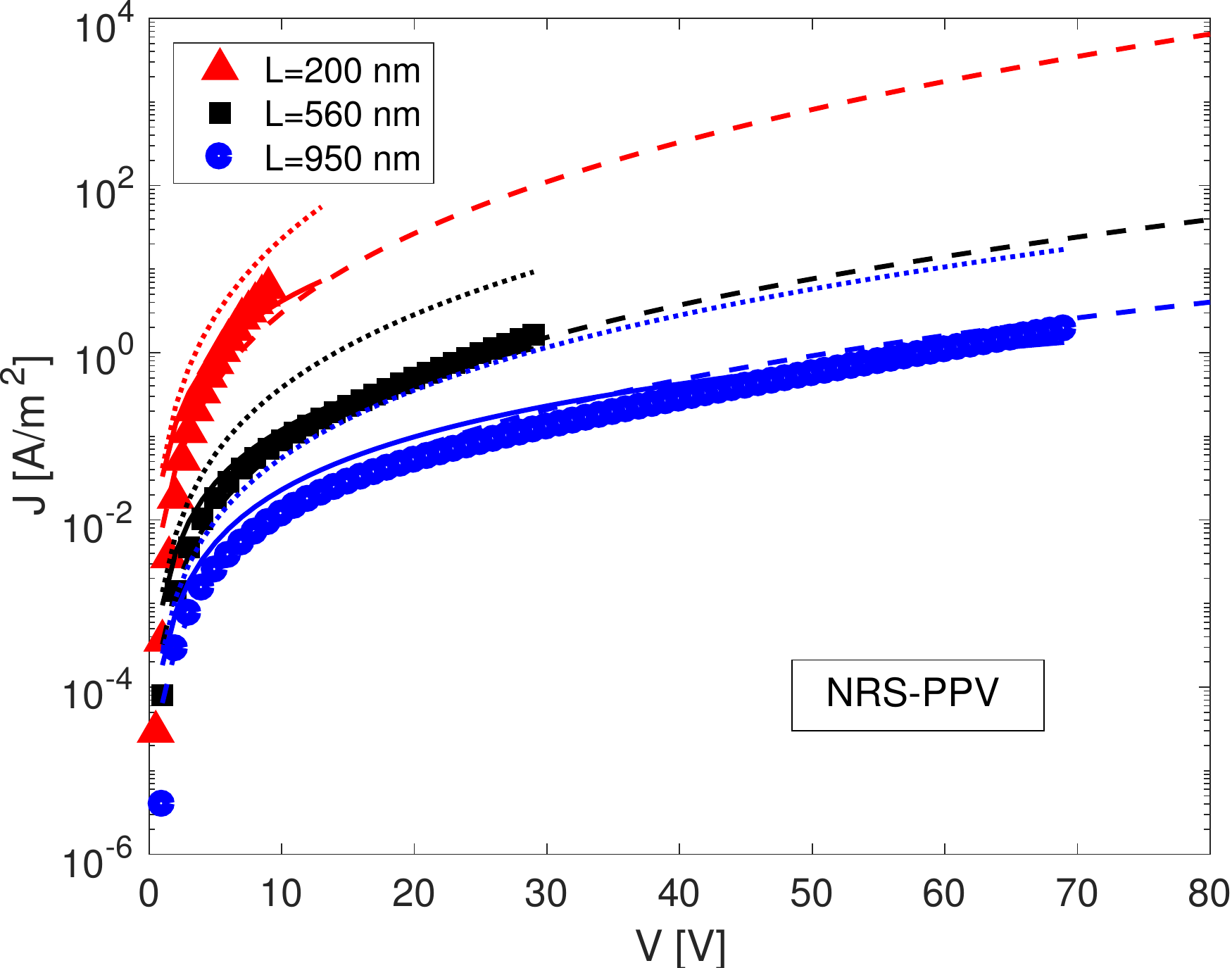}%
		\label{fig:APL2005} }
	\subfloat[\label{Fig:APL}]{%
		\includegraphics[width=.22\textwidth]{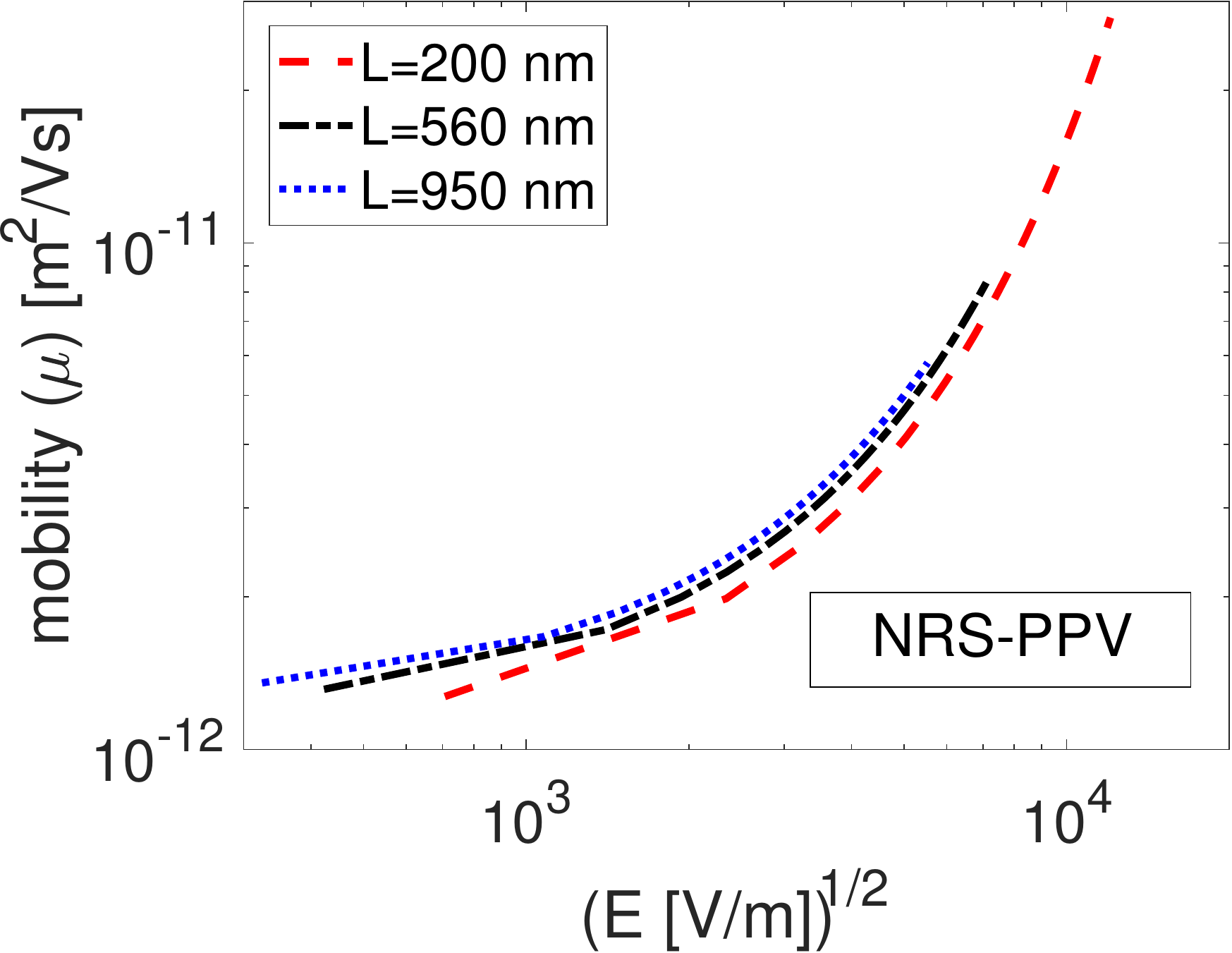}%
		\label{Fig:APL2005mobility}
	}\caption{\textbf{(a)} Room-temperature current density vs voltage characteristics data from~\cite{blom2005thickness} for NRS-PPV hole-only diodes with thicknesses of 200 (red), 560 (black), and 950 (blue) nm, respectively. Experiment (circles), Eq. \ref{eqn:MGClassical_fielddep} (dotted lines), Eq. \ref{eqn:MGClassicalFD_fielddep} (dashed lines), Eq. \ref{eqn:PowerLawFieldMobility-SCLC-FD} (solid lines). The parameters used are $\gamma=4\times10^{-4}(m/V)^{1/2}$, $\mu_{0}=5\times10^{-12} m^2/Vs$,  $\alpha=0.965$, $E_{0}=0.1/L^{\alpha}$, $n=0.10$. \textbf{(b)} The calculated mobility values for devices with varying thickness using $\mu=\mu_{0} \exp(0.89\gamma \sqrt{E}) L^{3-3\alpha}$.} \label{Fig:APL}
\end{figure}

In Fig. (\ref{fig:APL2005}) the experimental J-V
characteristics~\cite{blom2005thickness} (circles) of the NRS-PPV
based devices are shown together with various numerical models
calculations: (i) (dotted lines) classical MG law based
field-dependent mobility model of Eq.
(\ref{eqn:MGClassical_fielddep}), (ii) (dashed lines) modified MG
law based field-dependent mobility model of Eq.
(\ref{eqn:MGClassicalFD_fielddep}), and (iii) (solid lines) modified
MG law based field-dependent mobility model of Eq.
(\ref{eqn:PowerLawFieldMobility-SCLC-FD}). From the figure, it is
clear that the classical model (dotted lines) does not have a good
agreement with the experimental results. As shown in Fig.
(\ref{Fig:Slopes1}), NRS-PPV based devices show a thickness
dependence of $L^{-2.895}$ which corresponds to $\alpha=0.965$.
Using this $\alpha$ = 0.965, the two modified SCL models including
field dependent mobility (dashed and solid lines) are able to
provide better agreements without the needs to use carrier-dependent
mobility assumption that have been debated in recent
years~\cite{nenashev2015theoretical}. 
It is important to note that one of the direct consequence of modified MG law in Eq. (\ref{eqn:MGClassicalFD_fielddep}) is that the mobility can be considered to have a thickness dependence along with field-dependence given by $\mu=\mu_{0} \exp(0.89\gamma \sqrt{E}) L^{3-3\alpha}$. In Fig. (\ref{Fig:APL2005mobility}) we show the field and thickness dependent mobility values for the same NRS-PPV based devices~\cite{blom2005thickness} using this model with the parameters shown in figure caption.

\begin{figure}[!hb]
	\subfloat[\label{fig:RSC2014}]{%
		\includegraphics[width=.22\textwidth]{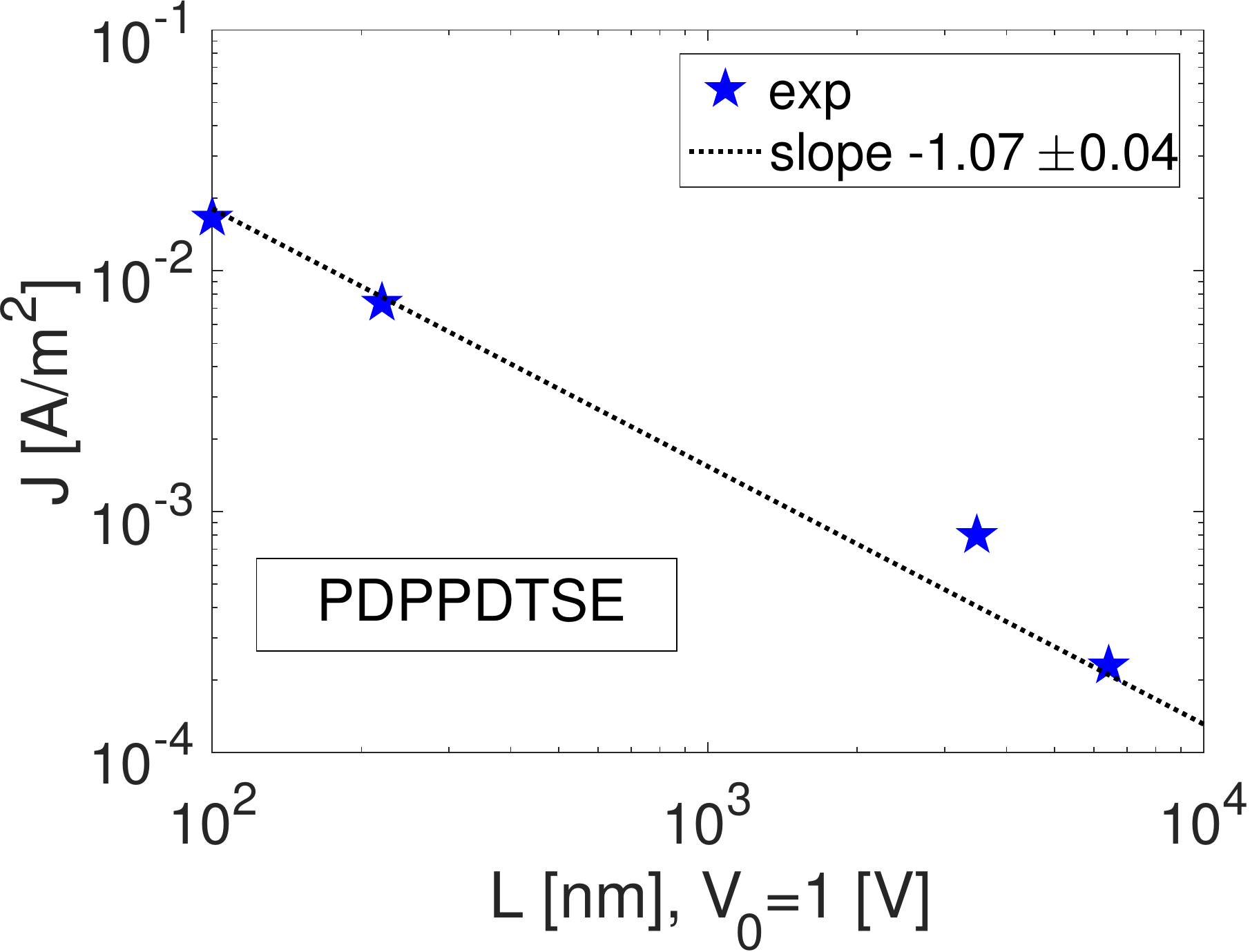}%
		\label{fig:RSC2014_J_vs_L} }
	\subfloat[\label{Fig:RSC2014b}]{%
		\includegraphics[width=.22\textwidth]{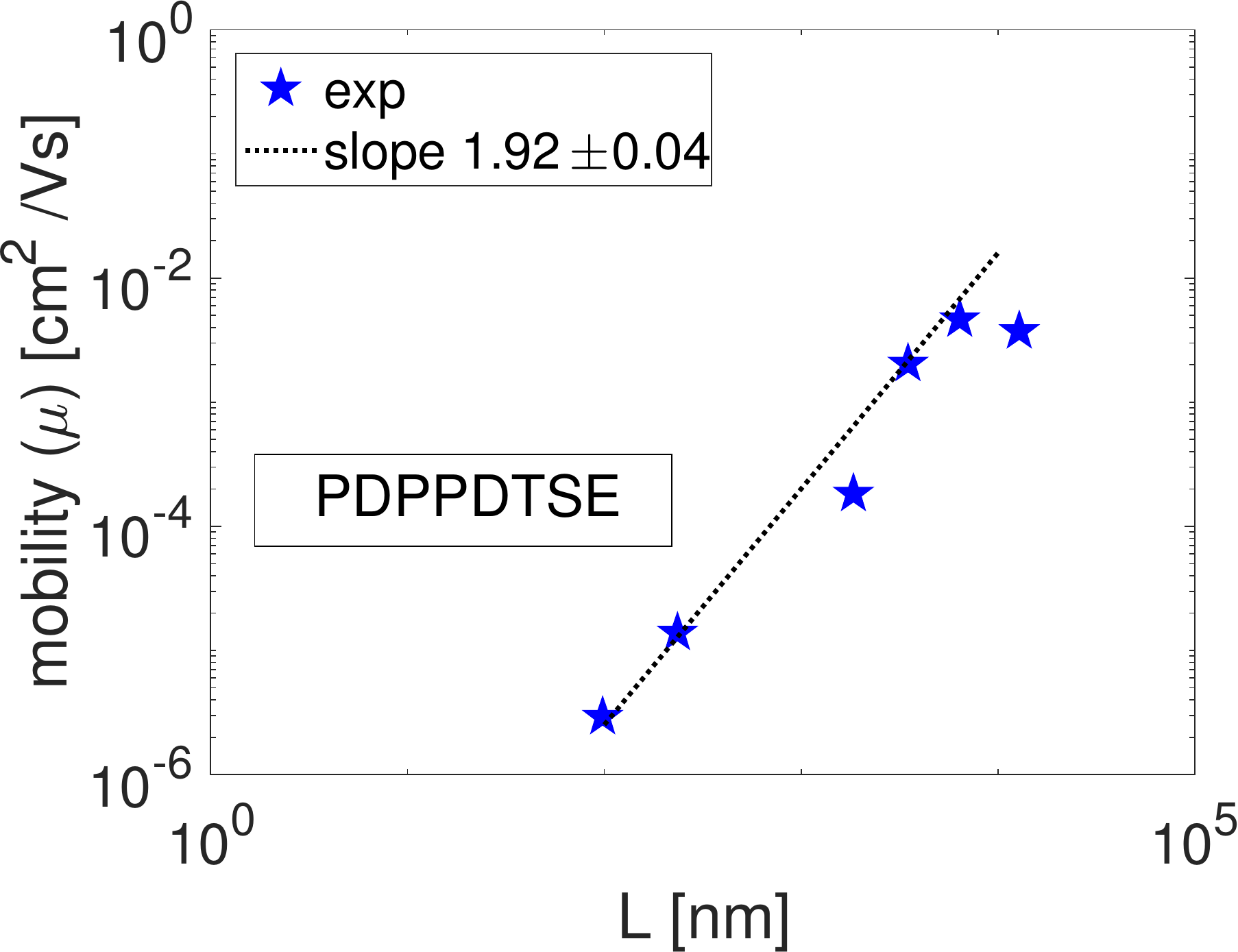}%
		\label{fig:RSC2014_L_vs_mobility}
	}\caption{\textbf{(a)} The thickness dependence of current density at fixed voltage for diketopyrrolopyrole-based polymer (PDPPDTSE) taken from experimental current-voltage data in the literature ~\cite{cheon2014analysis}. The observed thickness dependence corresponds to spatial disorder parameter $\alpha=0.3567$. \textbf{(b)} The thickness dependence of measured mobility agrees with the one predicted by our model ($L^{3-3\alpha}=L^{1.929}$).} \label{Fig:RSC2014}
\end{figure}

Finally, we analyzed the thickness dependence of experimentally measured SCLC in hole-only devices based on diketopyrrolopyrole-based polymer (PDPPDTSE)~\cite{cheon2014analysis}. The thickness dependence of current density at fixed voltage for PDPPDTSE based devices taken from experimental current-voltage data is shown in Fig. (\ref{fig:RSC2014_J_vs_L}). The thickness scaling of SCLC  from standard $L^{-3}$ to $L^{-1.07}$. The observed thickness dependence corresponds to spatial disorder parameter $\alpha=0.3567$. In order to validate our model we also compared the reported mobility values for varying thickness of devices with the mobility scaling predicted by our model. It is shown in Fig. (\ref{fig:RSC2014_L_vs_mobility})that the thickness scaling of measured mobility is in good agreement with the one predicted by our model ($L^{3-3\alpha}$).

\section{\label{sec: Summary} Summary}

\begin{table*}[!ht]
	\caption{Summary of modified SCLC models for spatially disordered organic semiconductors proposed in this work. See main text for complete description of terminologies.}
	\begin{center}
		\label{tab:summary}
		
		\begin{tabular}{lll}
			\hline
			\hline
			Description                                       & Modified SCLC Model                         & Eqs.    \\
			\hline
			\hline
			\begin{tabular}[c]{@{}l@{}}trap-free\\\\\end{tabular}                                                 & $J=
			\frac{9\alpha^{3}}{8}\left[\frac{\Gamma(\alpha/2)}{\pi^{\alpha/2}}\right]^3\epsilon
			\mu\frac{V_{0}^{2}}{L^{3\alpha}}$ & (13)  \\
			\begin{tabular}[c]{@{}l@{}}trap-limited \\ (exponential trap density)\\\\\end{tabular}&  $J= N_{c} \mu e^{1-l} \left[\frac{\Gamma(\alpha/2)}{\pi^{\alpha/2}}\right]^{2l+1} \left[ \frac{\epsilon \alpha l}{N_{t} (l+1)}\right]^{l} \nonumber\times\left( \frac{2 \alpha l+ \alpha}{l+1} \right)^{l+1} \frac{V_{0}^{l+1}}{L^{2 \alpha l+\alpha}}$& (21)\\
			\begin{tabular}[c]{@{}l@{}}trap-limited \\ (Gaussian trap density)\\\\\end{tabular}                                  & $J= N_{c} \mu e^{1-l}  \times\left[\frac{\Gamma(\alpha/2)}{\pi^{\alpha/2}}\right]^{2l+1} \left[ \frac{\epsilon \alpha l}{N_{t} \exp((E_{tc}-E_{a})/kT_c )(l+1)}\right]^{l} \times\left( \frac{2 \alpha l+ \alpha}{l+1} \right)^{l+1} \frac{V_{0}^{l+1}}{L^{2 \alpha l+\alpha}}$ & (23)\\
			\begin{tabular}[c]{@{}l@{}}field-dependent \\ (exponential)\\\\\end{tabular}                                         & $J=
			\frac{9\alpha^{3}}{8}\left[\frac{\Gamma(\alpha/2)}{\pi^{\alpha/2}}\right]^3\epsilon
			\mu_{0}\frac{V_{0}^{2}}{L^{3\alpha}} \exp(0.89 \gamma \sqrt{E})$ & (24) \\
			\begin{tabular}[c]{@{}l@{}}field-dependent \\ (power-law)\\\\\end{tabular}                                            & $J=
			\frac{\epsilon
				\mu_{0}}{E_{0}^{n}}\left[\frac{\Gamma(\alpha/2)}{\pi^{\alpha/2}}\right]^{n+3}
			\left[\frac{n\alpha-n+\alpha}{n+2}\right]\times
			\left[\frac{2n\alpha+3\alpha-n}{n+2}\right]^{n+2}
			\frac{V_{0}^{n+2}}{L^{2n\alpha+3\alpha-n}}$ & (26) \\
			
			\hline
			\hline
		\end{tabular}
	\end{center}
\end{table*}

In summary, we have presented a modified thickness scaling in SCLC model to account for
the spatial disorder in organic semiconductors by introducing a
parameter $\alpha$  to imagine the solid as a fractal object
sandwiched between two electrodes. 
The model has included different
effects such as trap-free, trap-limited and field-dependent mobility. To provide an easy access to the main results of this work, we have summarized the modified SCLC equations in Table II.
An analysis of multiple experimental results from literature reveals
that the classical SCLC models might lead to incorrect extraction of
mobilities due to weak thickness dependence arising from spatial
disorder. For such materials, our proposed model here would be a
better choice to extract the mobility for spatially disordered organic materials
as we have shown that the traditional thickness scaling is not valid
anymore. By applying our model with field-dependent mobility, we are
able to reproduce the experimental results of SCLC transport in PPV
derivative based device without using the carrier-density dependent
mobility~\cite{blom2005thickness}, agreeable with a recent report
for amorphous polymers~\cite{campbell2016charge}.

Note that the thickness dependence had been reported in others
works. For example, Brutting et. al (see Fig. 2a in Ref.
\cite{brutting2001space}) reported a weaker thickness dependence for
Alq light-emitting devices than the expected $L^{-1}$ at fixed
electric field. John et. al (see Fig.
(5-6) in Ref. \cite{john2010investigations}) reported a varying thickness dependence ($L^{-2.7\pm0.46}$ to
$L^{-3.14\pm0.7}$) for plasma polymerized pyrrole thin films. Boni
et. al. (see Fig. 12 in Ref. \cite{boni2013electronic}) also reported a
possible weaker thickness dependence for PZT ferroelectric based
devices. Macdonald et. al (see Fig. 1b in Ref.
\cite{macdonald2012electrical})  also reported a weaker thickness
dependence due to non-planar electrodes in conducting the experiment
using tip atomic force microscopy (cAFM). This is a geometrical
effect producing weaker thickness dependence of organic
semiconductor devices~\cite{reid2008space} and is different from the
physics studied here. It should be emphasized that our proposed
models are based on a planar-diode geometry, thus such non-planar
geometrical effects are not included. The extension of our models
into non-planar geometries will be pursued in future works. 

Moreover, in this work we obtain the parameter $\alpha$ from the length scaling of SCLC in the experimental results, however a complete microscopic model can be created in further extensions to determine $\alpha$ directly from the knowledge of disorder either spatial or energetic or both.

	

\appendices
\appendix

\section*{\label{sec:FDspace} Fractional-Dimensional Space Framework as Description of Complexity}

There is an increasing interest in the fractional modeling of
complexity in physical
systems~\cite{West2004fractional,tarasov2011fractional}. In recent years,
the concept of fractional-dimensional space has been used as an
effective physical description of restraint conditions in complex
physical
systems~\cite{palmer2004equations,tarasov2014anisotropic,zubair2012electromagnetic}.
The approaches to describe the fractional dimensions include fractal
geometry~\cite{falconer2004fractal}, fractional
calculus~\cite{oldham1974fractional,calcagni2012geometry}, and the
integration over fractional-dimensional
space~\cite{stillinger1977axiomatic,balankin2015effective}. The
axiomatic basis of spaces with fractional dimension with Euclidean metric were introduced by Stillinger~\cite{stillinger1977axiomatic}. The
fractional-dimensional generalization of first order Laplace
operators was then reported
by~Zubair et. al.~\cite{zubair2012electromagnetic} as approximations of the
square of the second-order Laplace operator introduced in~\cite{stillinger1977axiomatic,palmer2004equations}. Recently, a fractal metric based approach is considered by Tarasov~\cite{tarasov2014anisotropic} which provides a complete generalization of first and second order Laplace operators. In this work, we have utilized Tarasov's approach to vector calaculus in fractional-dimensional spaces, which is summarized in the following. 

In fractional-dimensional space ($F^{\alpha}
\subseteq E^{n}$) framework~\cite{tarasov2014anisotropic}, it is convenient to work with physically
dimensionless space variables $x/R_{0}\rightarrow x$,
$y/R_{0}\rightarrow y$, $z/R_{0}\rightarrow z$,
$\mathbf{r}/R_{0}\rightarrow \mathbf{r}$, where $R_{0}$ is a
characteristic size of considered model. This provides a
dimensionless integration and differentiation in
$\alpha$-dimensional space which leads to correct physical
dimensions of quantities.



We define a differential operator in the form of
\begin{eqnarray}
\partial_{\alpha_{k},x_{k}}=\frac{\partial}{\partial X_{k}}=\frac{1}{c(\alpha_{k},x_{k})}\frac{\partial}{\partial x_{k}},
\label{eqn:operatordiff}
\end{eqnarray}
where $c(\alpha_{k},x_{k})$ corresponds to the non-integer dimensionality along the $X_{k}$-axis and it is defined
by~\cite{ tarasov2014anisotropic}
\begin{eqnarray}
c(\alpha_{k},x_{k})=
\frac{\pi^{\alpha_{k}/2}}{\Gamma(\alpha_{k}/2)}|x_{k}|^{\alpha_{k}-1}.
\label{eqn:dos}
\end{eqnarray}

For the case of spatially disordered
semiconductor or porous solid, the system can be effectively modeled
by replacing the anisotropy with an isotropic continuum in an
$\alpha$-dimensional space,\ with a parameter 0 $<\alpha\leq$ 1  to
measure the anisotropy or disorder of the material.

Using the operators in Eq. (\ref{eqn:operatordiff}), we can generalize vector differential operators in an $\alpha$-dimensional space.
The gradient of a scalar function
$\varphi(\mathbf{r})$ in fractional-dimensional space is
\begin{eqnarray}
\nabla_{\alpha}\varphi(\mathbf{r})=\sum_{k=1}^{3}\mathbf{e}_{k}\partial_{\alpha_{k},x_{k}}\varphi(\mathbf{r}),
\label{eqn:grad}
\end{eqnarray}
where $\mathbf{e}_{k}$ are unit base vectors of the Cartesian
coordinate system. The divergence of the vector field
$\mathbf{f}(\mathbf{r})=\mathbf{e}_{k} f_{k} (\mathbf{r})$ is
\begin{eqnarray}
\nabla_{\alpha}\cdot\mathbf{f}(\mathbf{r})=\sum_{k=1}^{3}\partial_{\alpha_{k},x_{k}}f(\mathbf{r}).
\label{eqn:div}
\end{eqnarray}
The curl for the vector field $\mathbf{f}(\mathbf{r})$ is
\begin{eqnarray}
\nabla_{\alpha}\times\mathbf{f}(\mathbf{r})=\sum_{k,i,l=1}^{3}\mathbf{e}_{i}
\varepsilon _{ikl}\partial_{\alpha_{k},x_{k}}f(\mathbf{r}),
\label{eqn:curl}
\end{eqnarray}
where $\varepsilon _{ikl}$ is the Levi-Civita symbol.
Using Eqs. (\ref{eqn:grad}) and (\ref{eqn:div}), the scalar Laplacian in the fractional-dimensional-space is written as
\cite{tarasov2014anisotropic}
\begin{eqnarray}
\label{eqn:laplacian_scalar}
\nabla_{\alpha}^2
\varphi(\mathbf{r})&=&
\nabla_{\alpha}\cdot\nabla_{\alpha}\varphi(\mathbf{r})\\
\nonumber&=&\sum_{k=1}^{3}\frac{1}{c^2(\alpha_{k},x_{k})}\left(
\frac{\partial^2}{\partial
	x_{k}^2}-\frac{\alpha_{k}-1}{x_{k}}\frac{\partial}{\partial
	x_{k}}\right).
\end{eqnarray}

\bibliographystyle{IEEEtran}
\bibliography{DraftMGv3x1}

\providecommand{\noopsort}[1]{}\providecommand{\singleletter}[1]{#1}%
\begin{thebibliography}{10}
\providecommand{\url}[1]{#1}
\csname url@samestyle\endcsname
\providecommand{\newblock}{\relax}
\providecommand{\bibinfo}[2]{#2}
\providecommand{\BIBentrySTDinterwordspacing}{\spaceskip=0pt\relax}
\providecommand{\BIBentryALTinterwordstretchfactor}{4}
\providecommand{\BIBentryALTinterwordspacing}{\spaceskip=\fontdimen2\font plus
\BIBentryALTinterwordstretchfactor\fontdimen3\font minus
  \fontdimen4\font\relax}
\providecommand{\BIBforeignlanguage}[2]{{%
\expandafter\ifx\csname l@#1\endcsname\relax
\typeout{** WARNING: IEEEtran.bst: No hyphenation pattern has been}%
\typeout{** loaded for the language `#1'. Using the pattern for}%
\typeout{** the default language instead.}%
\else
\language=\csname l@#1\endcsname
\fi
#2}}
\providecommand{\BIBdecl}{\relax}
\BIBdecl

\bibitem{blom2005thickness}
P.~Blom, C.~Tanase, D.~De~Leeuw, and R.~Coehoorn, ``Thickness scaling of the
  space-charge-limited current in poly (p-phenylene vinylene),'' \emph{Applied
  physics letters}, vol.~86, no.~9, p. 092105, 2005.

\bibitem{kuik201425th}
M.~Kuik, G.-J.~A. Wetzelaer, H.~T. Nicolai, N.~I. Craciun, D.~M. De~Leeuw, and
  P.~W. Blom, ``25th anniversary article: Charge transport and recombination in
  polymer light-emitting diodes,'' \emph{Advanced Materials}, vol.~26, no.~4,
  pp. 512--531, 2014.

\bibitem{campbell2016charge}
A.~J. Campbell, R.~Rawcliffe, A.~Guite, J.~C.~D. Faria, A.~Mukherjee, M.~A.
  McLachlan, M.~Shkunov, and D.~D. Bradley, ``Charge-carrier density
  independent mobility in amorphous fluorene-triarylamine copolymers,''
  \emph{Advanced Functional Materials}, vol.~26, no.~21, pp. 3720--3729, 2016.

\bibitem{lampert1970current}
M.~A. Lampert and P.~Mark, ``Current injection in solids,'' 1970.

\bibitem{mott1948electronic}
N.~F. Mott and R.~W. Gurney, \emph{Electronic processes in ionic
  crystals}.\hskip 1em plus 0.5em minus 0.4em\relax Clarendon Press, 1948.

\bibitem{mark1962space}
P.~Mark and W.~Helfrich, ``Space-charge-limited currents in organic crystals,''
  \emph{Journal of Applied Physics}, vol.~33, no.~1, pp. 205--215, 1962.

\bibitem{murgatroyd1970theory}
P.~Murgatroyd, ``Theory of space-charge-limited current enhanced by frenkel
  effect,'' \emph{Journal of Physics D: Applied Physics}, vol.~3, no.~2, p.
  151, 1970.

\bibitem{bassler1993charge}
H.~B{\"a}ssler, ``Charge transport in disordered organic photoconductors a
  monte carlo simulation study,'' \emph{physica status solidi (b)}, vol. 175,
  no.~1, pp. 15--56, 1993.

\bibitem{blakesley2014towards}
J.~C. Blakesley, F.~A. Castro, W.~Kylberg, G.~F. Dibb, C.~Arantes, R.~Valaski,
  M.~Cremona, J.~S. Kim, and J.-S. Kim, ``Towards reliable charge-mobility
  benchmark measurements for organic semiconductors,'' \emph{Organic
  Electronics}, vol.~15, no.~6, pp. 1263--1272, 2014.

\bibitem{tanase2003unification}
C.~Tanase, E.~Meijer, P.~Blom, and D.~De~Leeuw, ``Unification of the hole
  transport in polymeric field-effect transistors and light-emitting diodes,''
  \emph{Physical Review Letters}, vol.~91, no.~21, p. 216601, 2003.

\bibitem{fishchuk2007analytic}
I.~Fishchuk, V.~Arkhipov, A.~Kadashchuk, P.~Heremans, and H.~B{\"a}ssler,
  ``Analytic model of hopping mobility at large charge carrier concentrations
  in disordered organic semiconductors: Polarons versus bare charge carriers,''
  \emph{Physical Review B}, vol.~76, no.~4, p. 045210, 2007.

\bibitem{pasveer2005unified}
W.~Pasveer, J.~Cottaar, C.~Tanase, R.~Coehoorn, P.~Bobbert, P.~Blom,
  D.~De~Leeuw, and M.~Michels, ``Unified description of charge-carrier
  mobilities in disordered semiconducting polymers,'' \emph{Physical review
  letters}, vol.~94, no.~20, p. 206601, 2005.

\bibitem{cottaar2011scaling}
J.~Cottaar, L.~Koster, R.~Coehoorn, and P.~Bobbert, ``Scaling theory for
  percolative charge transport in disordered molecular semiconductors,''
  \emph{Physical review letters}, vol. 107, no.~13, p. 136601, 2011.

\bibitem{tanase2004charge}
C.~Tanase, P.~Blom, D.~De~Leeuw, and E.~Meijer, ``Charge carrier density
  dependence of the hole mobility in poly (p-phenylene vinylene),''
  \emph{physica status solidi (a)}, vol. 201, no.~6, pp. 1236--1245, 2004.

\bibitem{fishchuk2010temperature}
I.~Fishchuk, A.~Kadashchuk, J.~Genoe, M.~Ullah, H.~Sitter, T.~B. Singh,
  N.~Sariciftci, and H.~B{\"a}ssler, ``Temperature dependence of the charge
  carrier mobility in disordered organic semiconductors at large carrier
  concentrations,'' \emph{Physical Review B}, vol.~81, no.~4, p. 045202, 2010.

\bibitem{katsouras2013charge}
I.~Katsouras, A.~Najafi, K.~Asadi, A.~Kronemeijer, A.~Oostra, L.~Koster, D.~M.
  de~Leeuw, and P.~W. Blom, ``Charge transport in poly (p-phenylene vinylene)
  at low temperature and high electric field,'' \emph{Organic Electronics},
  vol.~14, no.~6, pp. 1591--1596, 2013.

\bibitem{leijtens2013charge}
T.~Leijtens, J.~Lim, J.~Teuscher, T.~Park, and H.~J. Snaith, ``Charge density
  dependent mobility of organic hole-transporters and mesoporous tio2
  determined by transient mobility spectroscopy: Implications to dye-sensitized
  and organic solar cells,'' \emph{Advanced Materials}, vol.~25, no.~23, pp.
  3227--3233, 2013.

\bibitem{nenashev2015theoretical}
A.~Nenashev, J.~Oelerich, and S.~Baranovskii, ``Theoretical tools for the
  description of charge transport in disordered organic semiconductors,''
  \emph{Journal of Physics: Condensed Matter}, vol.~27, no.~9, p. 093201, 2015.

\bibitem{tessler2009charge}
N.~Tessler, Y.~Preezant, N.~Rappaport, and Y.~Roichman, ``Charge transport in
  disordered organic materials and its relevance to thin-film devices: A
  tutorial review,'' \emph{Advanced Materials}, vol.~21, no.~27, pp.
  2741--2761, 2009.

\bibitem{noriega2013general}
R.~Noriega, J.~Rivnay, K.~Vandewal, F.~P. Koch, N.~Stingelin, P.~Smith, M.~F.
  Toney, and A.~Salleo, ``A general relationship between disorder, aggregation
  and charge transport in conjugated polymers,'' \emph{Nature materials},
  vol.~12, no.~11, pp. 1038--1044, 2013.

\bibitem{zubair2017fractional}
M.~Zubair, Y.~S. Ang, and L.~K. Ang, ``Fractional fowler-nordheim law for field
  emission from rough surface with nonparabolic energy dispersion,'' \emph{IEEE
  Transactions on Electron Devices}, vol.~65, no.~6, pp. 2089--2095, 2018.

\bibitem{zubair2016fractional}
M.~Zubair and L.~K. Ang, ``Fractional-dimensional child-langmuir law for a
  rough cathode,'' \emph{Physics of Plasmas (1994-present)}, vol.~23, no.~7, p.
  072118, 2016.

\bibitem{stillinger1977axiomatic}
F.~H. Stillinger, ``Axiomatic basis for spaces with noninteger dimension,''
  \emph{Journal of Mathematical Physics}, vol.~18, no.~6, pp. 1224--1234, 1977.

\bibitem{palmer2004equations}
C.~Palmer and P.~N. Stavrinou, ``Equations of motion in a
  non-integer-dimensional space,'' \emph{Journal of Physics A: Mathematical and
  General}, vol.~37, no.~27, p. 6987, 2004.

\bibitem{sadallah2009solution}
M.~Sadallah and S.~I. Muslih, ``Solution of the equations of motion for
  einstein's field in fractional d dimensional space-time,''
  \emph{International Journal of Theoretical Physics}, vol.~48, no.~12, pp.
  3312--3318, 2009.

\bibitem{tarasov2016heat}
V.~E. Tarasov, ``Heat transfer in fractal materials,'' \emph{International
  Journal of Heat and Mass Transfer}, vol.~93, pp. 427--430, 2016.

\bibitem{ostoja2014fractal}
M.~Ostoja-Starzewski, J.~Li, H.~Joumaa, and P.~N. Demmie, ``From fractal media
  to continuum mechanics,'' \emph{ZAMM-Journal of Applied Mathematics and
  Mechanics/Zeitschrift f{\"u}r Angewandte Mathematik und Mechanik}, vol.~94,
  no.~5, pp. 373--401, 2014.

\bibitem{balankin2012map}
A.~S. Balankin and B.~E. Elizarraraz, ``Map of fluid flow in fractal porous
  medium into fractal continuum flow,'' \emph{Physical Review E}, vol.~85,
  no.~5, p. 056314, 2012.

\bibitem{zubair2012electromagnetic}
M.~Zubair, M.~J. Mughal, and Q.~A. Naqvi, \emph{Electromagnetic fields and
  waves in fractional dimensional space}.\hskip 1em plus 0.5em minus
  0.4em\relax Springer Science \& Business Media, 2012.

\bibitem{mughal2011fractional}
M.~J. Mughal and M.~Zubair, ``Fractional space solutions of antenna radiation
  problems: An application to hertzian dipole,'' in \emph{Signal Processing and
  Communications Applications (SIU), 2011 IEEE 19th Conference on}.\hskip 1em
  plus 0.5em minus 0.4em\relax IEEE, 2011, pp. 62--65.

\bibitem{naqvi2016cylindrical}
Q.~A. Naqvi and M.~Zubair, ``On cylindrical model of electrostatic potential in
  fractional dimensional space,'' \emph{Optik-International Journal for Light
  and Electron Optics}, vol. 127, no.~6, pp. 3243--3247, 2016.

\bibitem{zubair2011exact}
M.~Zubair, M.~J. Mughal, and Q.~A. Naqvi, ``An exact solution of the
  cylindrical wave equation for electromagnetic field in fractional dimensional
  space,'' \emph{Progress In Electromagnetics Research}, vol. 114, pp.
  443--455, 2011.

\bibitem{asad2012electromagnetic}
H.~Asad, M.~J. Mughal, M.~Zubair, and Q.~A. Naqvi, ``Electromagnetic green's
  function for fractional space,'' \emph{Journal of Electromagnetic Waves and
  Applications}, vol.~26, no. 14-15, pp. 1903--1910, 2012.

\bibitem{asad2012reflection}
H.~Asad, M.~Zubair, and M.~J. Mughal, ``Reflection and transmission at
  dielectric-fractal interface,'' \emph{Progress In Electromagnetics Research},
  vol. 125, pp. 543--558, 2012.

\bibitem{zubair2011exact2}
M.~Zubair, M.~J. Mughal, and Q.~A. Naqvi, ``An exact solution of the spherical
  wave equation in d-dimensional fractional space,'' \emph{Journal of
  Electromagnetic Waves and Applications}, vol.~25, no.~10, pp. 1481--1491,
  2011.

\bibitem{zubair2011differential}
M.~Zubair, M.~J. Mughal, Q.~A. Naqvi, and A.~A. Rizvi, ``Differential
  electromagnetic equations in fractional space,'' \emph{Progress In
  Electromagnetics Research}, vol. 114, pp. 255--269, 2011.

\bibitem{zubair2011electromagnetic}
M.~Zubair, M.~J. Mughal, and Q.~A. Naqvi, ``On electromagnetic wave propagation
  in fractional space,'' \emph{Nonlinear Analysis: Real World Applications},
  vol.~12, no.~5, pp. 2844--2850, 2011.

\bibitem{zubair2010wave}
M.~\vspace{0mm}Zubair, M.~J. Mughal, and Q.~A. Naqvi, ``The wave equation and
  general plane wave solutions in fractional space,'' \emph{Progress In
  Electromagnetics Research Letters}, vol.~19, pp. 137--146, 2010.

\bibitem{tarasov2014anisotropic}
V.~E. Tarasov, ``Anisotropic fractal media by vector calculus in non-integer
  dimensional space,'' \emph{Journal of Mathematical Physics}, vol.~55, no.~8,
  p. 083510, 2014.

\bibitem{hwang1976studies}
W.~Hwang and K.~Kao, ``Studies of the theory of single and double injections in
  solids with a gaussian trap distribution,'' \emph{Solid-State Electronics},
  vol.~19, no.~12, pp. 1045--1047, 1976.

\bibitem{abbaszadeh2016elimination}
D.~Abbaszadeh, A.~Kunz, G.~Wetzelaer, J.~Michels, N.~Cra?ciun, K.~Koynov,
  I.~Lieberwirth, and P.~Blom, ``Elimination of charge carrier trapping in
  diluted semiconductors,'' \emph{Nature materials}, vol.~15, no.~6, pp.
  628--633, 2016.

\bibitem{mandoc2007trap}
M.~Mandoc, B.~de~Boer, G.~Paasch, and P.~Blom, ``Trap-limited electron
  transport in disordered semiconducting polymers,'' \emph{Physical Review B},
  vol.~75, no.~19, p. 193202, 2007.

\bibitem{chen1978model}
I.~Chen, ``A model of charge injection at metal-insulator contacts,''
  \emph{Solid State Communications}, vol.~26, no.~6, pp. 359--363, 1978.

\bibitem{abkowitz1993time}
M.~Abkowitz, J.~Facci, and M.~Stolka, ``Time-resolved space charge-limited
  injection in a trap-free glassy polymer,'' \emph{Chemical physics}, vol. 177,
  no.~3, pp. 783--792, 1993.

\bibitem{nicolai2010space}
H.~Nicolai, G.~Wetzelaer, M.~Kuik, A.~Kronemeijer, B.~De~Boer, and P.~Blom,
  ``Space-charge-limited hole current in poly (9, 9-dioctylfluorene) diodes,''
  \emph{Applied Physics Letters}, vol.~96, no.~17, p. 172107, 2010.

\bibitem{horowitz1990evidence}
G.~Horowitz, D.~Fichou, X.~Peng, and P.~Delannoy, ``Evidence for a linear
  low-voltage space-charge-limited current in organic thin films. film
  thickness and temperature dependence in alpha-conjugated sexithienyl,''
  \emph{Journal de Physique}, vol.~51, no.~13, pp. 1489--1499, 1990.

\bibitem{coehoorn2006measurement}
R.~Coehoorn, S.~Vulto, S.~Van~Mensfoort, J.~Billen, M.~Bartyzel, H.~Greiner,
  and R.~Assent, ``Measurement and modeling of carrier transport and exciton
  formation in blue polymer light emitting diodes,'' in \emph{Photonics
  Europe}.\hskip 1em plus 0.5em minus 0.4em\relax International Society for
  Optics and Photonics, 2006, pp. 61\,920O--61\,920O.

\bibitem{mandoc2006electron}
M.~Mandoc, B.~De~Boer, and P.~Blom, ``Electron-only diodes of poly
  (dialkoxy-p-phenylene vinylene) using hole-blocking bottom electrodes,''
  \emph{Physical Review B}, vol.~73, no.~15, p. 155205, 2006.

\bibitem{chu2007hole}
T.-Y. Chu and O.-K. Song, ``Hole mobility of n, n'-bis (naphthalen-1-yl)-n,
  n'-bis (phenyl) benzidine investigated by using space-charge-limited
  currents,'' \emph{Applied physics letters}, vol.~90, no.~20, pp.
  203\,512--203\,512, 2007.

\bibitem{cheon2014analysis}
K.~H. Cheon, J.~Cho, B.~T. Lim, H.-J. Yun, S.-K. Kwon, Y.-H. Kim, and D.~S.
  Chung, ``Analysis of charge transport in high-mobility diketopyrrolopyrole
  polymers by space charge limited current and time of flight methods,''
  \emph{RSC Advances}, vol.~4, no.~67, pp. 35\,344--35\,347, 2014.

\bibitem{brutting2001space}
W.~Br{\"u}tting, S.~Berleb, and A.~M{\"u}ckl, ``Space-charge limited conduction
  with a field and temperature dependent mobility in alq light-emitting
  devices,'' \emph{Synthetic Metals}, vol. 122, no.~1, pp. 99--104, 2001.

\bibitem{john2010investigations}
J.~John, S.~Sivaraman, S.~Jayalekshmy, and M.~Anantharaman, ``Investigations on
  the mechanism of carrier transport in plasma polymerized pyrrole thin
  films,'' \emph{Journal of Physics and Chemistry of Solids}, vol.~71, no.~7,
  pp. 935--939, 2010.

\bibitem{boni2013electronic}
A.~Boni, I.~Pintilie, L.~Pintilie, D.~Preziosi, H.~Deniz, and M.~Alexe,
  ``Electronic transport in (la, sr) mno3-ferroelectric-(la, sr) mno3 epitaxial
  structures,'' \emph{Journal of Applied Physics}, vol. 113, no.~22, p. 224103,
  2013.

\bibitem{macdonald2012electrical}
G.~A. MacDonald, P.~A. Veneman, D.~Placencia, and N.~R. Armstrong, ``Electrical
  property heterogeneity at transparent conductive oxide/organic semiconductor
  interfaces: mapping contact ohmicity using conducting-tip atomic force
  microscopy,'' \emph{ACS nano}, vol.~6, no.~11, pp. 9623--9636, 2012.

\bibitem{reid2008space}
O.~G. Reid, K.~Munechika, and D.~S. Ginger, ``Space charge limited current
  measurements on conjugated polymer films using conductive atomic force
  microscopy,'' \emph{Nano letters}, vol.~8, no.~6, pp. 1602--1609, 2008.

\bibitem{West2004fractional}
B.~J. West, \emph{Fractional calculus view of complexity: Tomorrow's
  science}.\hskip 1em plus 0.5em minus 0.4em\relax CRC Press, 2015.

\bibitem{tarasov2011fractional}
V.~E. Tarasov, \emph{Fractional dynamics: applications of fractional calculus
  to dynamics of particles, fields and media}.\hskip 1em plus 0.5em minus
  0.4em\relax Springer Science \& Business Media, 2011.

\bibitem{falconer2004fractal}
K.~Falconer, \emph{Fractal geometry: mathematical foundations and
  applications}.\hskip 1em plus 0.5em minus 0.4em\relax John Wiley \& Sons,
  2004.

\bibitem{oldham1974fractional}
K.~B. Oldham and J.~Spanier, \emph{The Fractional Calculus}.\hskip 1em plus
  0.5em minus 0.4em\relax Academic Press, New York, 1974.

\bibitem{calcagni2012geometry}
G.~Calcagni, ``Geometry and field theory in multi-fractional spacetime,''
  \emph{Journal of High Energy Physics}, vol. 2012, no.~1, pp. 1--77, 2012.

\bibitem{balankin2015effective}
A.~S. Balankin, ``Effective degrees of freedom of a random walk on a fractal,''
  \emph{Physical Review E}, vol.~92, no.~6, p. 062146, 2015.

\end{thebibliography}

\begin{IEEEbiography}[{\includegraphics[width=1in,height=1.25in,clip,keepaspectratio]{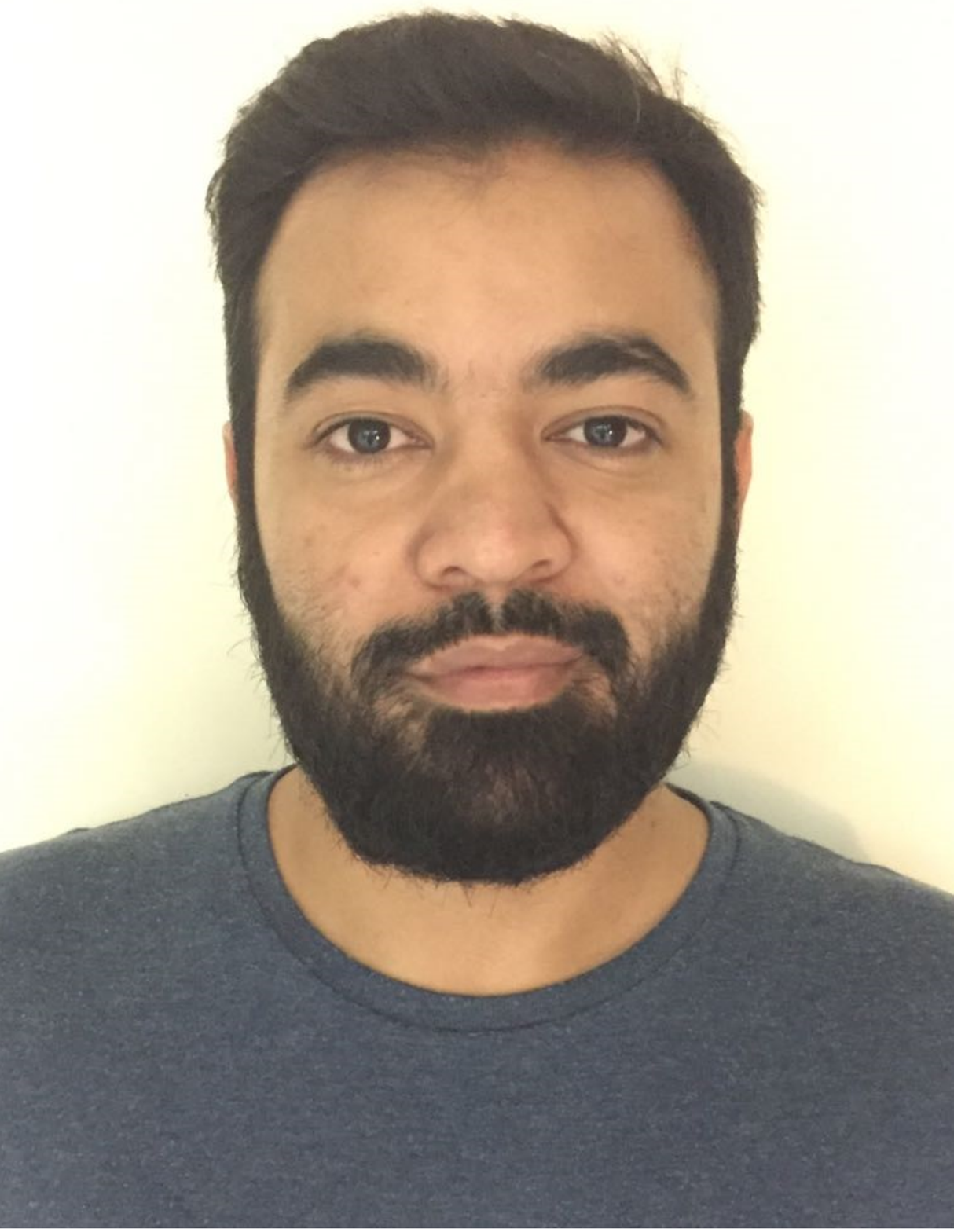}}]{Muhammad Zubair} (S'13-M'15) received his Ph.D. degree in electronic engineering from the Politecnico di Torino, Italy, in 2015. From 2015 to 2017, he was with the SUTD-MIT International Design Center, Singapore. Since 2017, he has been with Information Technology University, Lahore, Pakistan. 
	
	His current research interests include charge transport, electron device modeling, computational electromagnetics, fractal electrodynamics, and microwave imaging.
\end{IEEEbiography}
\begin{IEEEbiography}[{\includegraphics[width=1in,height=1.25in,clip,keepaspectratio]{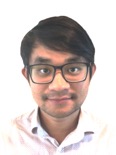}}]{Yee Sin Ang} his bachelor’s degree in medical and radiation physics in 2010, and his PhD degree in theoretical condensed matter physics in 2014 from the University of Wollongong (UOW), Australia. He is currently a Research Fellow with the Singapore University of Technology and Design, Singapore. 
	
	His research interests include the theory and mathematical modelling of electron emission phenomena in 2D and topological materials, electron transport physics across 2D/3D, 2D material valleytronics, nanoelectronics and superconducting devices. 
	
\end{IEEEbiography}
\begin{IEEEbiography}[{\includegraphics[width=1in,height=1.25in,clip,keepaspectratio]{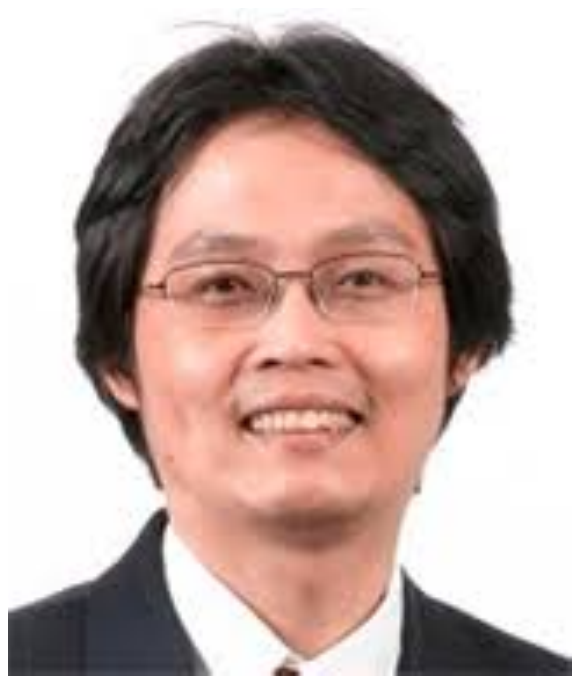}}]{Lay Kee Ang } (S'95-M'00-SM'08) received the B.S. degree from the Department of Nuclear Engineering, National Tsing Hua University, Hsinchu, Taiwan, in 1994, and the M.S. and Ph.D. degrees from the Department of Nuclear Engineering and Radiological Sciences, University of Michigan, Ann Arbor, MI, USA, in 1996 and 1999, respectively. Since 2011, he has been with the Singapore University of Technology and Design, Singapore. 
	
	He is currently the Interim Head and Professor of the Engineering Product Development pillar and also the Ng Teng Fong Chair Professor of SUTD-ZJU IDEA.
\end{IEEEbiography}

\end{document}